\documentclass[preprint,showpacs,preprintnumbers,amsmath,amssymb,floatfix]{revtex4-1}

\usepackage{graphicx,color}
\usepackage{dcolumn}
\usepackage{bm}
\usepackage{supertabular}

\begin{document}

\title{Improvement of low energy atmospheric neutrino flux calculation using 
the JAM nuclear interaction model}
\author{M.~Honda}
\email[]{mhonda@icrr.u-tokyo.ac.jp}
\homepage[]{http://icrr.u-tokyo.ac.jp/~mhonda}
\affiliation{Institute for Cosmic Ray Research, the University of Tokyo, 5-1-5 Kashiwa-no-ha, Kashiwa, Chiba 277-8582, Japan}
\author{T.~Kajita}
\email[]{kajita@icrr.u-tokyo.ac.jp}
\affiliation{Institute for Cosmic Ray Research, and
Institute for the Physics and Mathematical of the Universe, the University of Tokyo, 5-1-5 Kashiwa-no-ha, Kashiwa, Chiba 277-8582, Japan }
\author{K.~Kasahara}
\email[]{kasahara@icrc.u-tokyo.ac.jp}
\affiliation{Research Institute for Science and Engineering, Waseda University, 3-4-1 Okubo Shinjuku-ku, Tokyo, 169-8555, Japan.}
\author{S.~Midorikawa}
\email[]{midori@aomori-u.ac.jp}
\affiliation{Faculty of Software and Information Technology, Aomori University, Aomori, 030-0943 Japan.}
\date{\today}

\begin{abstract}

We present the calculation of the atmospheric neutrino fluxes with an 
interaction model named JAM,
which is used in PHITS (Particle and Heavy-Ion Transport code 
System)~\cite{phits}.
The JAM interaction model agrees with the HARP experiment~\cite{harp:p-air}
a little better than DPMJET-III~\cite{dpm}. 
After some modifications, 
it reproduces the muon flux below 1~GeV$/$c at balloon altitudes 
better than the modified-DPMJET-III which we used for the calculation of 
atmospheric neutrino flux in previous works~\cite{shkkm2006,hkkms2006}.
Some improvements in the calculation of atmospheric neutrino flux 
are also reported.
\end{abstract}

\maketitle

\section{introduction}

We have calculated atmospheric neutrino fluxes using a 3D-scheme,
and refined it, in previous works~\cite{hkkm2004,hkkms2006}.
In Ref~\cite{hkkm2004} 
we studied reputable interaction models with the
atmospheric muon spectrum at balloon altitudes, and
selected DPMJET-III~\cite{dpm}.
We then constructed and used an inclusive interaction code,
which has virtually the same secondary production 
spectra as DPMJET-III, but which runs far more rapidly
(see appendix~\ref{inclusive-code}).
However, when we compare the high energy muon flux calculated using 
DPMJET-III with the observed fluxes at sea level and mountain altitudes,
there are obvious deficiencies in the calculated values at 
high momenta ($\gtrsim 30$~GeV$/$c)~\cite{shkkm2006}.
Therefore, we modified the ``inclusive DPMJET-III'' so that the calculation
reproduces the observed atmospheric muon spectra with better accuracy
at high momenta, and used it in the calculation of the atmospheric neutrino 
fluxes~\cite{hkkms2006}. 
We called the modified inclusive interaction code the 
``modified DPMJET-III''.
Below 5~GeV we used the inclusive NUCRIN which is constructed from
NUCRIN~\cite{nucrin} with some modifications.
However, the contributions of the hadronic interaction below 5~GeV 
to the atmospheric neutrino and muon fluxes even at around 1~GeV$/$c and below
are smaller than those above 5~GeV.

With the ``modified DPMJET-III'', we could reproduce the observed 
atmospheric muon spectra quite well above 1~GeV$/$c at sea level
and mountain altitudes,
but the agreement below 1 GeV$/$c is not as good.
At balloon altitudes in particular the calculated muon fluxes are 
obviously lower than the observed ones.
This seems to be due to the secondary spectra of DPMJET-III,
which our simple modification procedure
(see appendix~\ref{inclusive-code}) can not correct.
The disagreement was not resolved without violating the 
agreement at higher momenta.
Instead of more sophisticated modification method,
we searched for an interaction models which has a better nature at low energies,
and found the interaction model called JAM, which is used
in PHITS (Particle and Heavy-Ion Transport code System)~\cite{phits}.
JAM is useful from much lower projectile energies ($<<$ 0.1 GeV) 
up to $\sim$ 100~GeV.

Apart from the atmospheric muon flux, 
the HARP experiment has greatly improved our knowledge of the hadronic interaction 
at low energies~\cite{harp:p-c}.
It is notable that the hadronic interactions of protons on thin N$_2$ 
and O$_2$ targets have been studied in detail by the HARP experiment, aiming 
to improve the calculation of atmospheric neutrino fluxes~\cite{harp:p-air}.
They found that the commonly used hadronic interaction models including
DPMJET-III do not reproduce the result of HARP experiment 
accurately, especially for the $\pi^-$ productions~\cite{harp:p-c}.
We examine JAM also with the HARP data, and find 
the JAM agrees a little better than DPMJET-III.

Then, we construct the inclusive interaction code for JAM
and studied the atmospheric muon flux with it at sea level, at mountain
altitudes, and at balloon altitudes.
The ``inclusive JAM'' is used for hadronic interactions 
from 0.2\,GeV to 32\,GeV, and above these energies, the
modified DPMJET-III is used as before.
After some modifications for the ``inclusive JAM'', 
we can reproduce the observed atmospheric muon flux below 1\,GeV$/$c at
sea level and at balloon altitudes, but
the agreement of the atmospheric muon flux at the mountain altitudes becomes
worse.
As we can carry out a study similar to the previous 
publication~\cite{shkkm2006} between the muons at balloon altitudes
and the interaction model relevant to the low energy ($\lesssim$~1~GeV)
atmospheric neutrinos,
we consider the agreement at balloon altitudes to be more important,
and use it for the calculation of the atmospheric neutrino fluxes.

The calculation scheme is essentially the same 
as that used in Ref~\cite{hkkm2004},
although we are refining it in each work.
In this work, we optimized the ``virtual detector correction'' to 
minimize the statistical errors.
Also with the increase of the available computation power, we 
calculate the atmospheric neutrino fluxes in this 3D scheme up to 32~GeV.
We find that
the calculated atmospheric neutrino fluxes show a visible 
azimuthal dependence below 32~GeV.  

The change of interaction model and refinement of 
the calculation scheme notwithstanding,
the zenith angle dependence of the atmospheric neutrino fluxes and the 
the $(\nu_\mu + \bar\nu_\mu)/(\nu_e + \bar\nu_e)$ ratio are very close to
the previous calculations.
The change of the hadronic interaction model mainly results in 
a change to the absolute 
value of the atmospheric neutrino flux below 1\,GeV.

\section{\label{intmodel}Interaction model and calculation of atmospheric neutrino flux}

The requirements for the hadronic interaction model used in the calculation of 
atmospheric neutrino fluxes are that it reproduce the real hadronic interaction
accurately,
and that the code generates hadronic events very quickly.
The first requirement is a general requirement for all applications of the 
hadronic interaction model.

We collected the reputable interaction models, and examined them with 
available accelerator data.
We further examined them with the atmospheric muon flux observed at balloon 
altitudes.
As a result, we selected DPMJET-III for our calculations in 
Ref~\cite{hkkm2004}.
Note, DPMJET-III is not usable below $\sim$5~GeV in projectile
energy, and so we used NUCRIN below 5~GeV.

The second requirement often conflicts with the first one.
The computer code of a sophisticated interaction model generally 
takes a lot of time to generate an interaction event.
To resolve this conflict,
we construct a quick inclusive interaction code, which generates virtually 
the same secondary production spectra as the original interaction model 
from the output of the original code.
We explain the construction of the inclusive code in some detail 
in appendix~\ref{inclusive-code}.
The interaction code used in Ref~\cite{hkkm2004} was 
such an inclusive code, made from the output of original code of 
DPMJET-III, and was called the ``inclusive DPMJET-III''.
The inclusive code is useful for the calculation of the time
averaged quantity, but is not usable for the simulation of an event
caused by a high energy cosmic ray, such as an air shower.

\begin{figure*}[!t]
  \centering
  \includegraphics[width=5in]{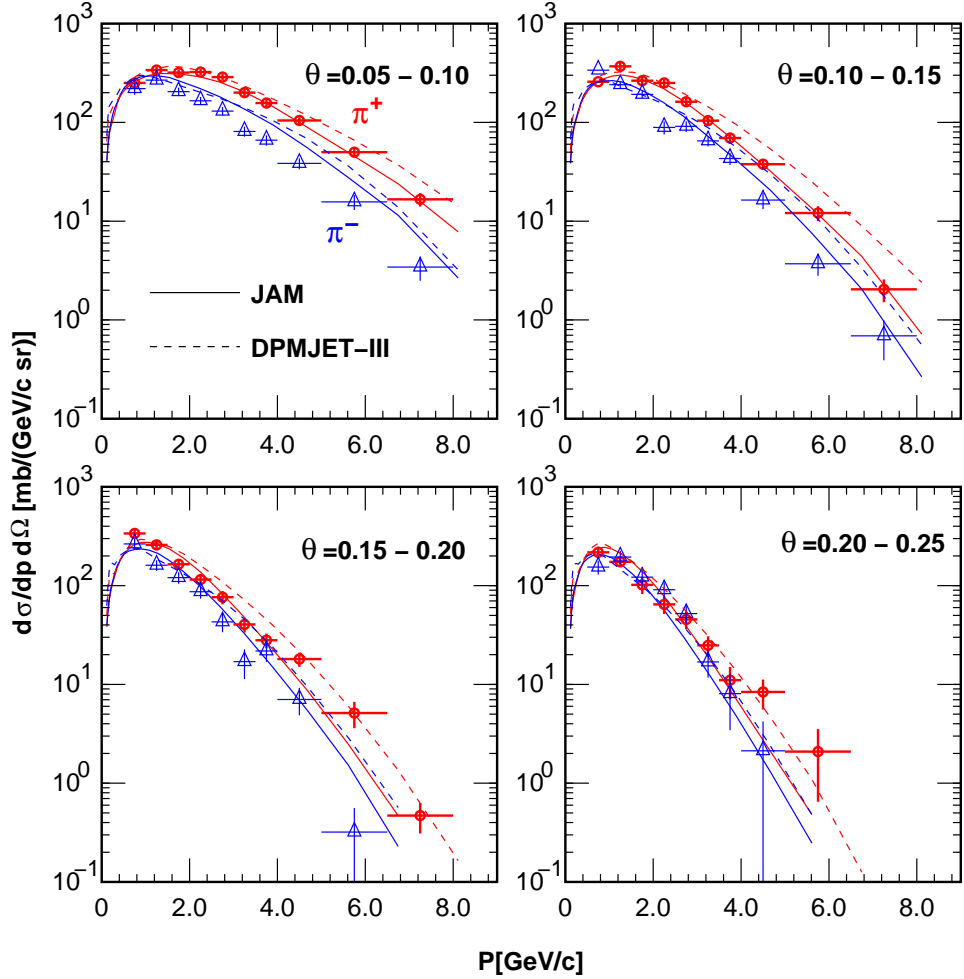}
  \caption{Comparison of interaction models and HARP data.
The solid lines stand for JAM, and the dashed lines for DPMJET-III.}
  \label{Harp-comp-4}
 \end{figure*}

The use of the inclusive code has the advantage that we can simply modify the
secondary spectra of the original interaction model without detailed knowledge
of the assumptions internal to it.
We modified the inclusive DPMJET-III  
in the procedure explained in appendix~\ref{inclusive-code}, 
and
we could reproduce the observed atmospheric muon flux above 1\,GeV
with satisfactory accuracy at sea-level and mountain altitude
\cite{shkkm2006}.
The modification changes the gradient of secondary spectra globally,
and is effectively a change of the so called $Z$-factor ($Z \equiv <x^{1.7}>$ 
with $x \equiv E_{secondary}/E_{projectile}$),
which describes the atmospheric muon and neutrino fluxes at high energies well
\cite{Gaisser-semi-analytic}.
The power of 1.7 is approximately the spectral index of the integral 
cosmic ray nucleon flux.
We called it modified-DPMJET-III, and 
calculated the atmospheric neutrino flux with it~\cite{hkkms2006}.

With the modification, however, we could not reproduce the observed 
atmospheric muon flux below $\sim$1\,GeV$/$c.
The cosmic ray energies relevant to the atmospheric muon at 
1~GeV$/$c are around 10~GeV or lower, depending on the observing 
altitude,
and the cosmic ray spectrum can no longer be approximated by a single 
power law there.
The structure of the secondary spectra, which does not have much 
influence on the $Z$-factor, becomes important.
The disagreement between calculations and observations in the low energy 
atmospheric muon is due to such a structure in the DPMJET-III model
at low projectile energies ($\lesssim 10$~GeV).
One approach to improving the agreement would be
developing a more sophisticated and complicated modification 
procedure.
However, we have taken an alternative approach 
and looked for an interaction model which has a better low energy 
behavior than DPMJET-III,
and have found an interaction model called JAM
which is used in the PHITS (Particle and Heavy-Ion Transport code 
System)\cite{phits}.

\begin{figure}[htb]
  \centering{
  \includegraphics[height=2.5in]{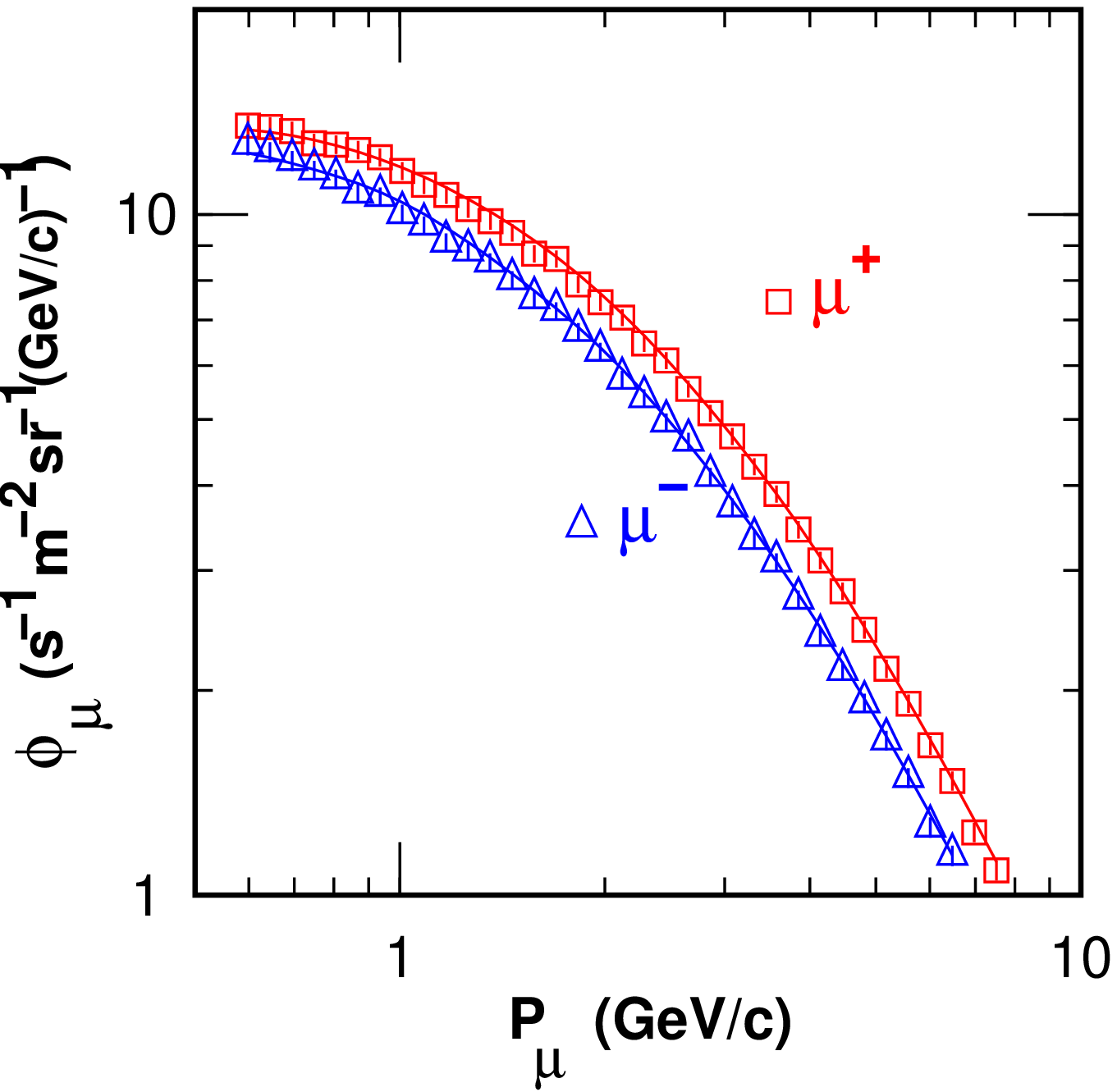}
  \includegraphics[height=2.5in]{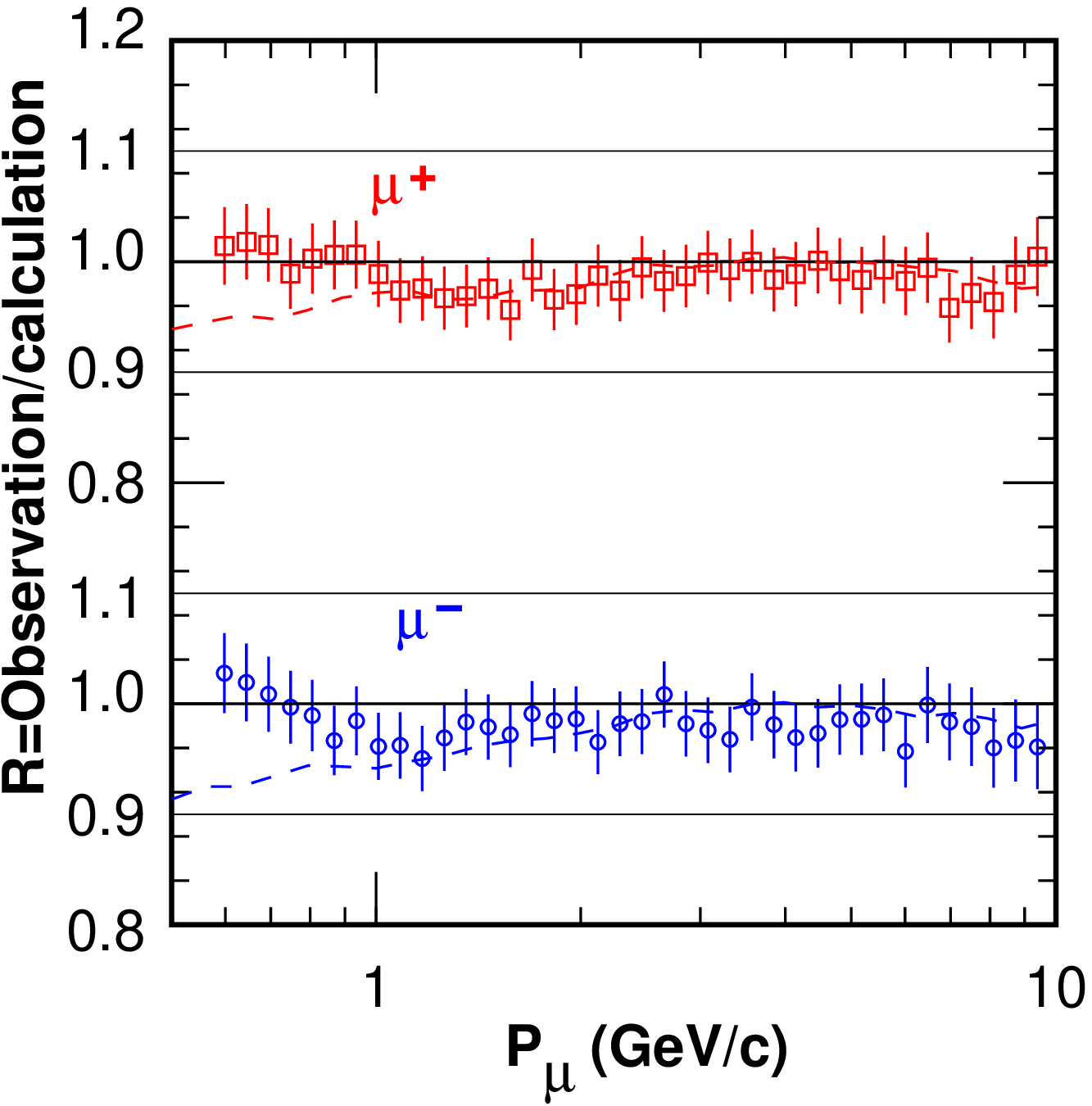}
  }\caption{Comparison of the muon flux observed at Tsukuba (30\,m a.s.l.)
and the one calculated with the modified-JAM. 
Dashed lines in right panel are the ratios of 
former calculations to present ones.}
  \label{test-tsukuba}
\end{figure}

After our previous calculation, the HARP collaboration released 
a much more detailed study of hadronic interactions of protons on  
thin N$_2$ and O$_2$ targets at 12\,GeV/c~\cite{harp:p-air}.
We compared the HARP data with the 
outputs of DPMJET-III and JAM in Fig.\ref{Harp-comp-4}.
In the figure, we combine N$_2$ and O$_2$ target data as the data for 
air target,
assuming the composition of air to be 78.5\% N$_2$ and 21.5\% O$_2$.
We also assume the 
inelastic cross section for proton and air nuclei interactions to be 300\,mb.
In the figure, we find that JAM shows slightly better agreement with 
the HARP experiment than DPMJET-III.
Also the secondary spectra of JAM is higher than that of DPMJET-III
at lower energies.

\begin{figure}[htb]
  \centering{
  \includegraphics[height=2.5in]{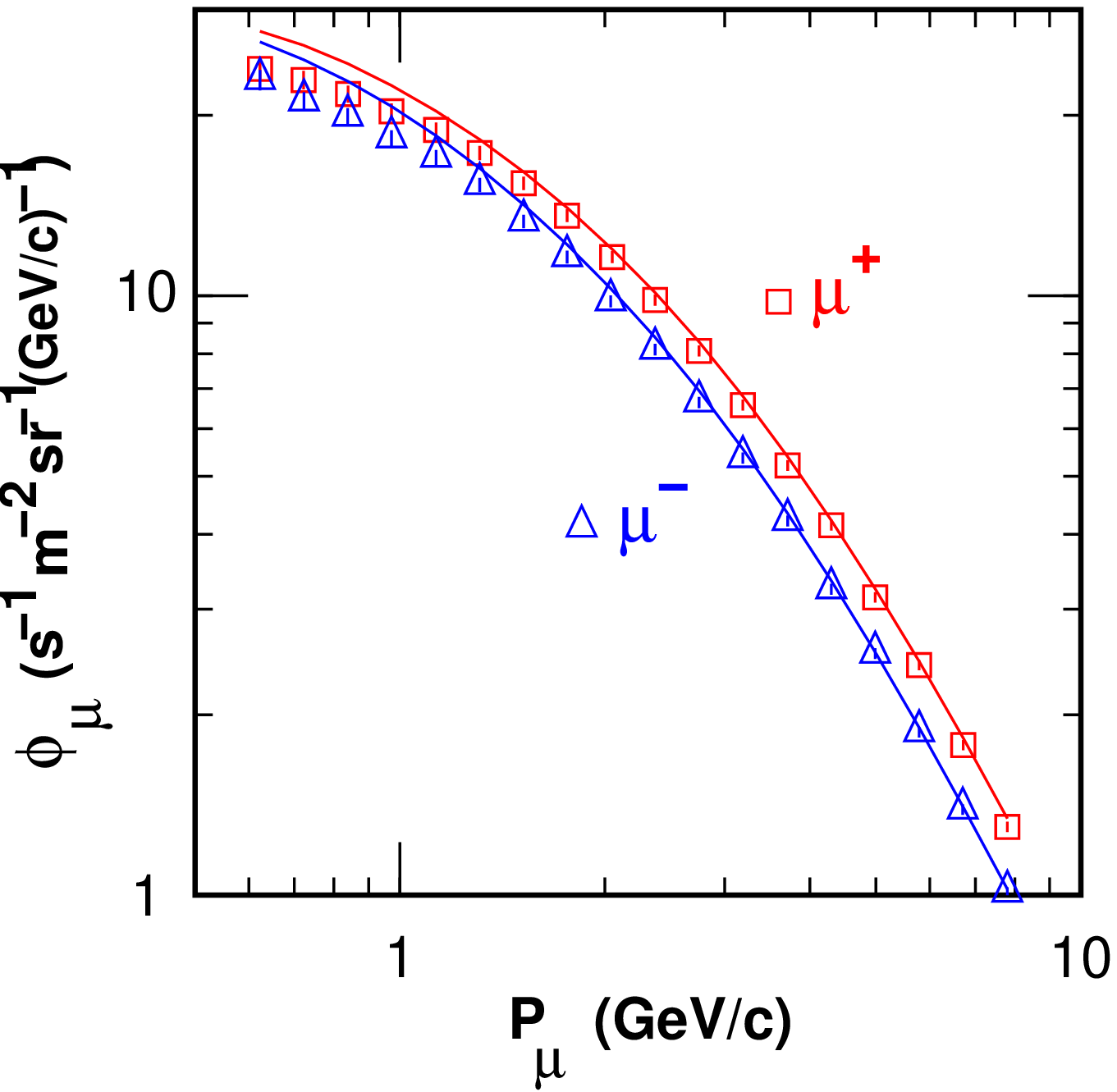}
  \includegraphics[height=2.5in]{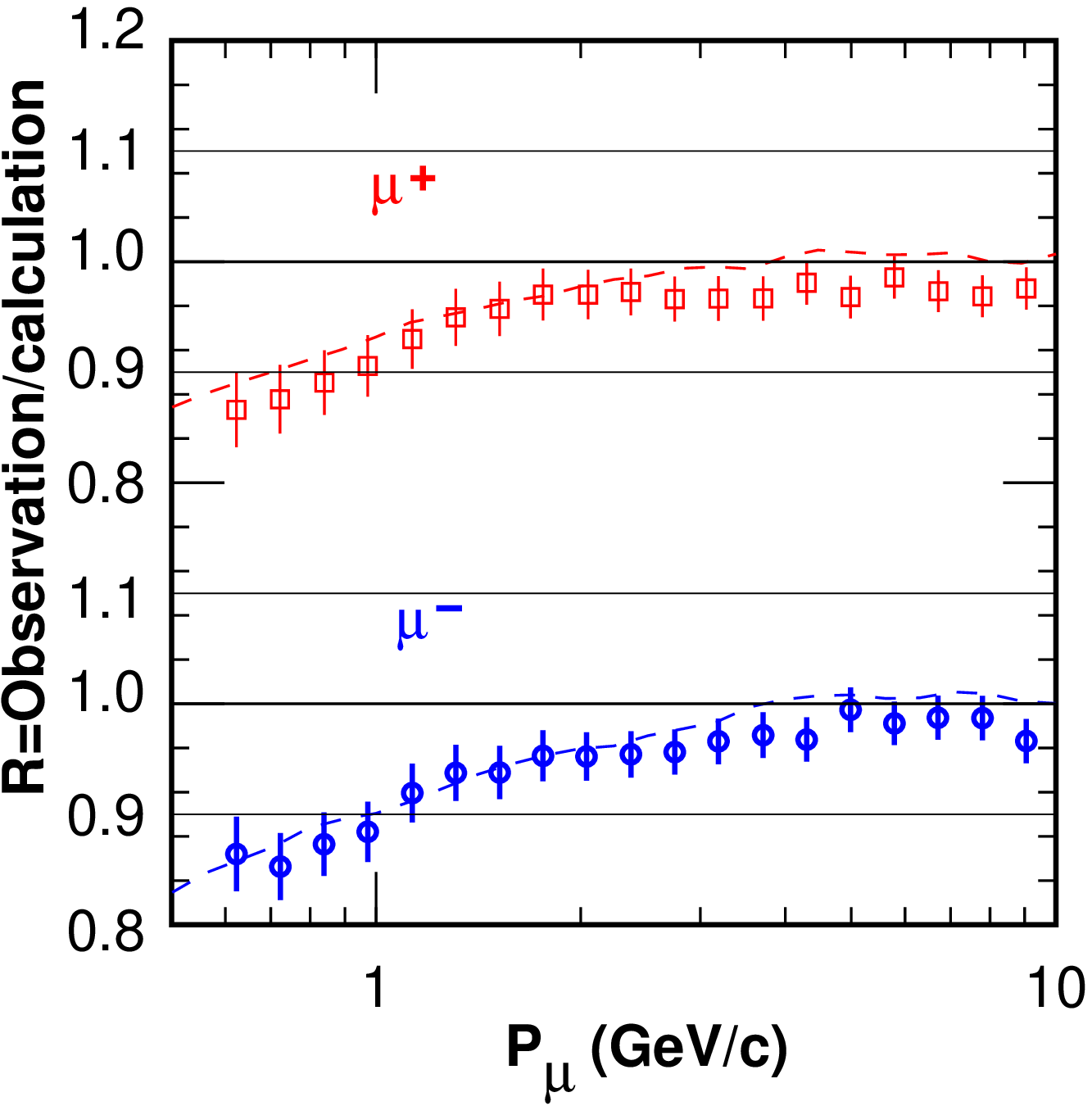}
  }\caption{Comparison of the muon flux observed at Mt.\ Norikura (2770\,m a.s.l)
and the one calculated with the modified-JAM.
Dashed lines in right panel are the ratios of 
former calculations to present ones.}
  \label{test-norikura}
\end{figure}

Constructing the inclusive JAM,
we calculate the atmospheric muon flux at the three different altitudes  
where the BESS group has carried out experimental studies of the atmospheric 
muon fluxes, at Tsukuba (30\,m a.s.l.)~\cite{BESSTeVpHemu}, 
at Mt.\ Norikura (2770\,m a.s.l.)~\cite{BESSnorimu}, and at Fort Sumner
(balloon altitudes)~\cite{Abe:2003cd}.
As the original JAM seems to have an upper limit for the projectile 
energy ($\lesssim$ 100 GeV), we use it below 32\,GeV.
Above that energy we use the inclusive DPMJET-III,
as the modified DPMJET-III reproduces the observed muon 
flux data above $\gtrsim$1\, GeV/c.

\begin{figure}[htb]
  \centering{
  \includegraphics[height=2.5in]{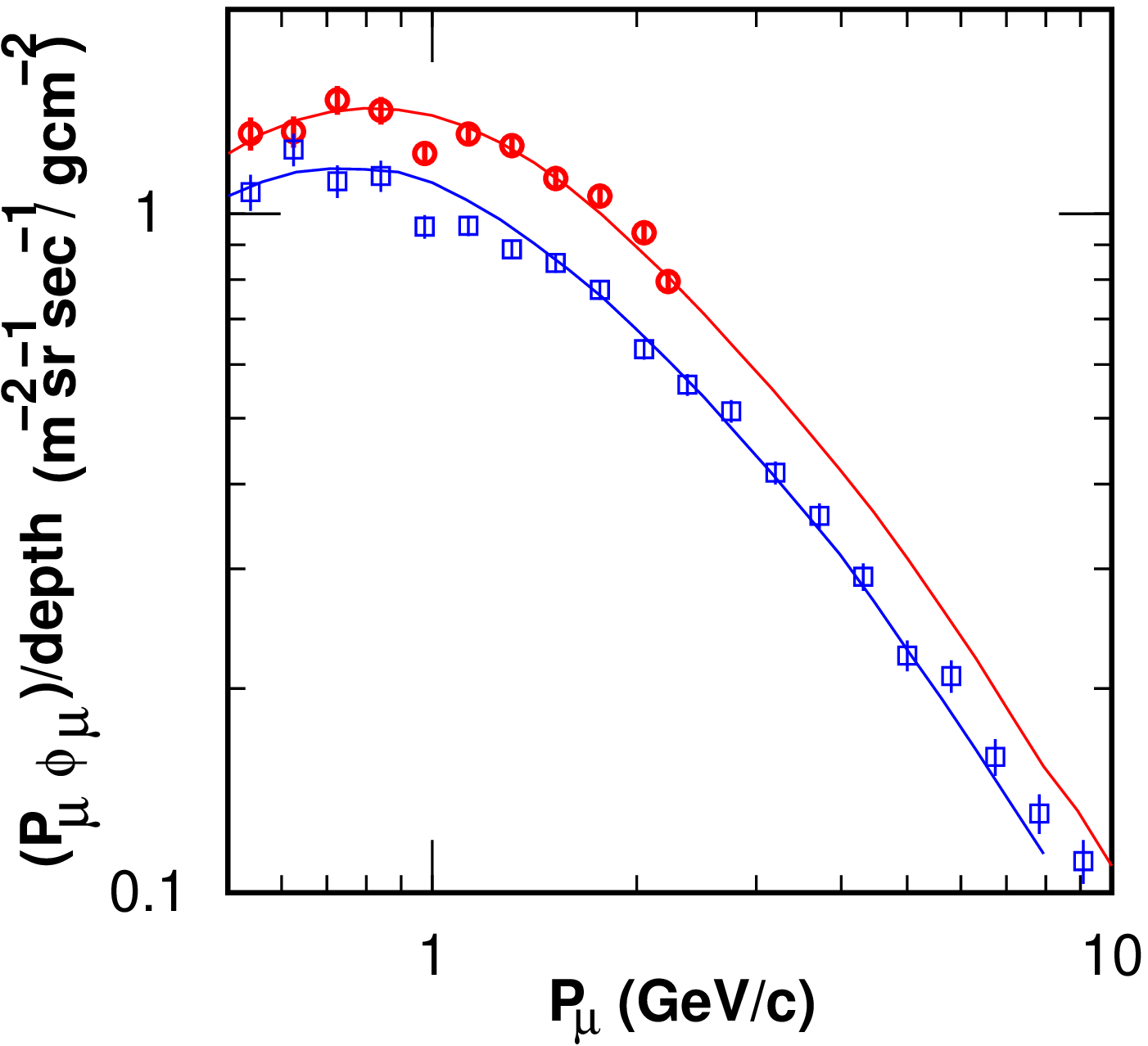}
  \includegraphics[height=2.5in]{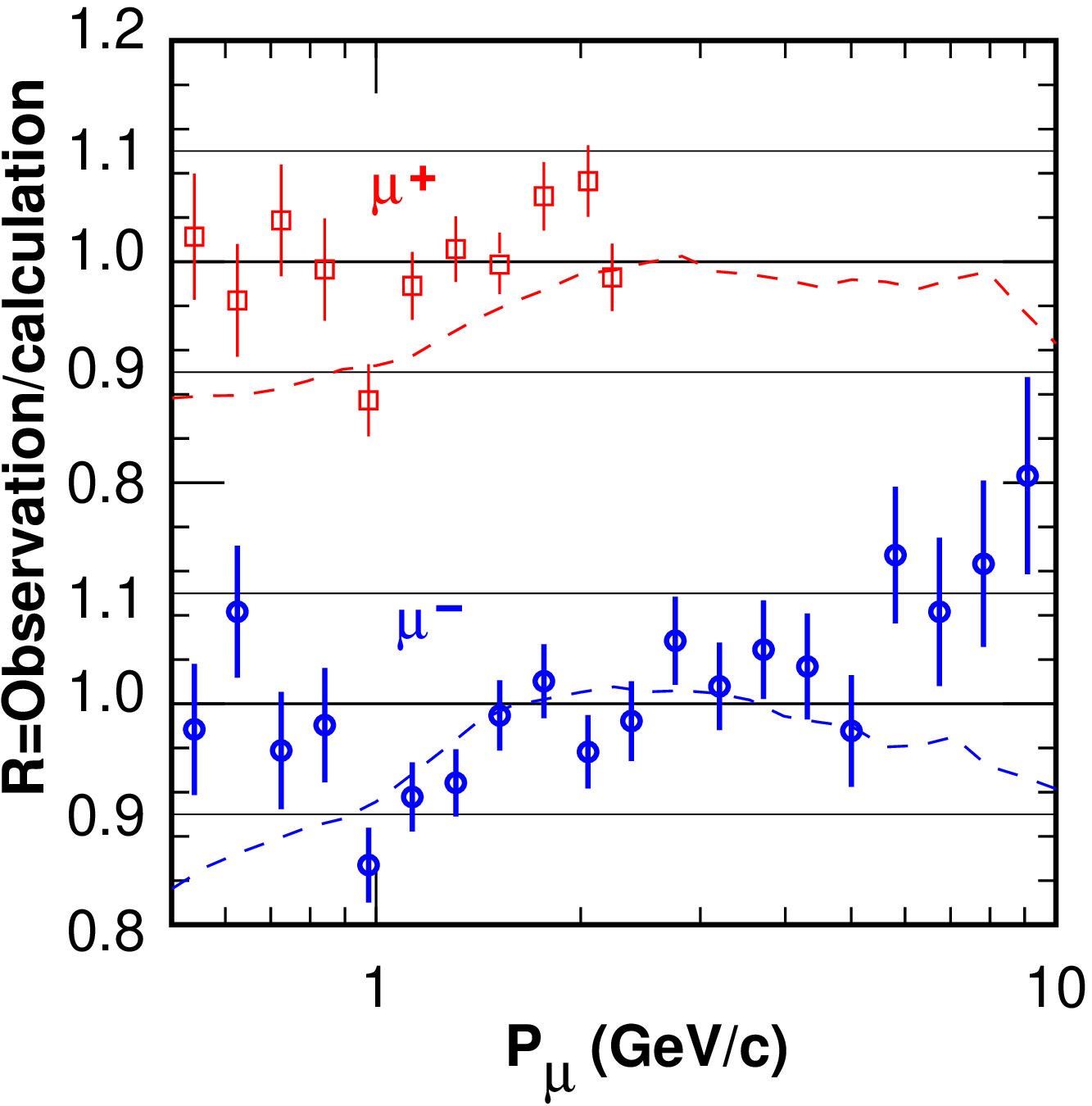}
 } \caption{Comparison of the muon flux observed at the balloon altitudes 
(5--2\,g$/$cm$^2$)
at Fort Sumner and the one calculated with the modified-JAM.
Dashed lines in right panel are the ratios of 
former calculations to present ones.}
  \label{test-fort}
\end{figure}

We show the comparisons of the calculations using the modified JAM 
with observations in  
Figs.~\ref{test-tsukuba}, \ref{test-norikura}, and \ref{test-fort},
as the ratios of the observed muon fluxes to the calculated ones.
In these calculations, we used the air density profile measured on 
the same date as the observations. This makes the agreement between the calculation
and observation better for $\gtrsim$1\,GeV$/$c.
Some modifications are also applied to the JAM model, following the procedure 
explained in appendix~\ref{inclusive-code}.

We note that
it was very difficult to modify the inclusive DPMJET-III only 
to give the muon flux 
as large as the observed one at Tsukuba and at Fort Sumner observed by BESS 
below 1\,GeV$/$c.
Introducing the inclusive JAM with the some modifications,
we can boost the calculated muon flux below 1\,GeV,
and get good agreement between the calculations and observations at Tsukuba 
and Fort Sumner.
However, boosting the muon flux below 1\,GeV makes the 
difference between the calculation and observations at Mt.\ Norikura larger.
We consider that it is more important to reduce the difference at balloon 
altitudes,
since the atmospheric muon at the balloon altitude is directly related
to the primary cosmic rays and the hadronic interactions at the energy 
relevant to the low energy ($\lesssim$~1~GeV) atmospheric neutrinos.
We can carry out a study of the hadronic interaction at this energy
as we did in the 
the previous publication~\cite{shkkm2006} using the muons observed at 
ground level at the energy relevant to the atmospheric neutrino 
with energies above 1\,GeV.
We calculate the atmospheric neutrino flux with the 
modified JAM in this paper.

\begin{figure}[htb]
  \centering{
  \includegraphics[height=2.2in]{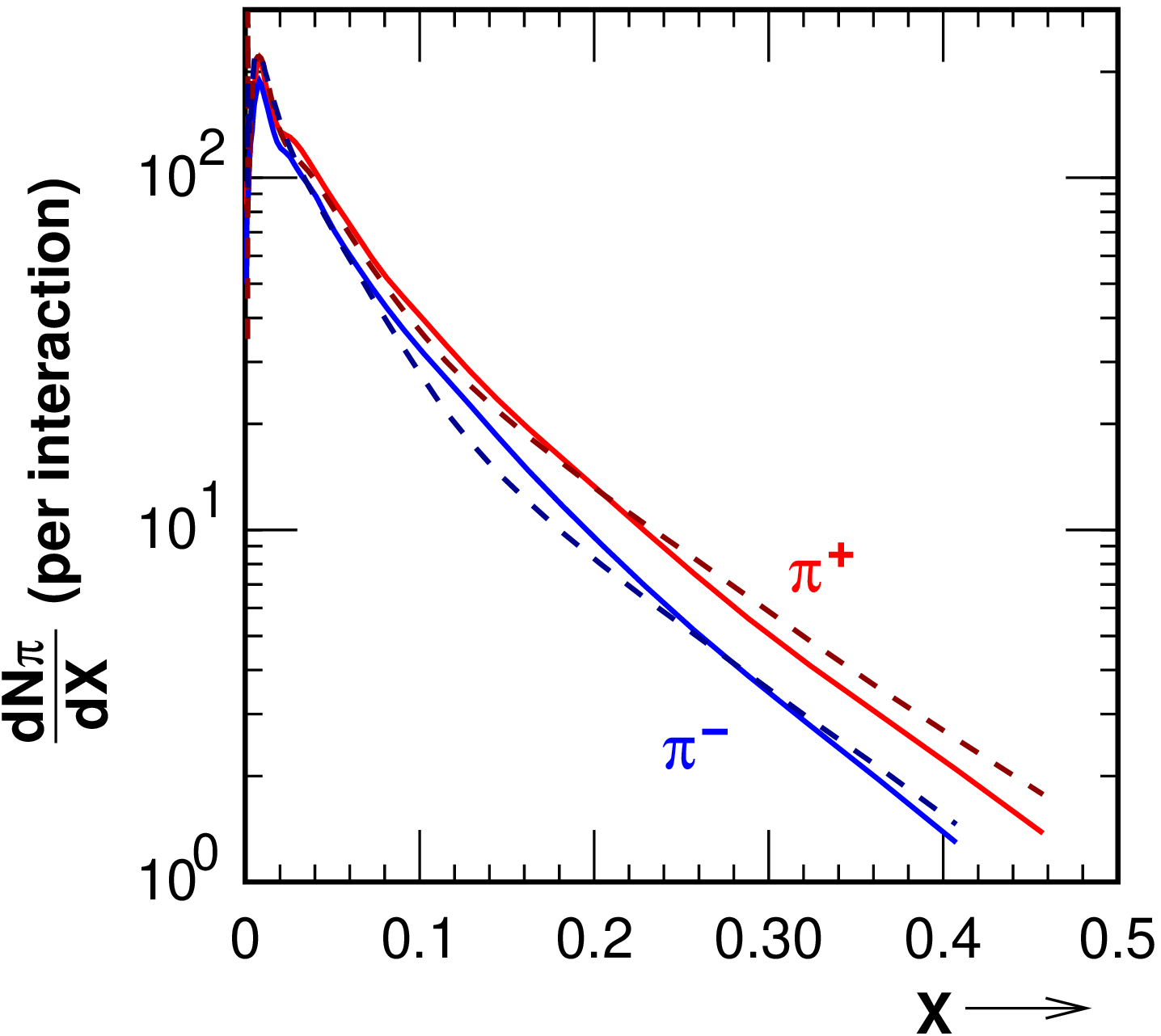}
  \includegraphics[height=2.2in]{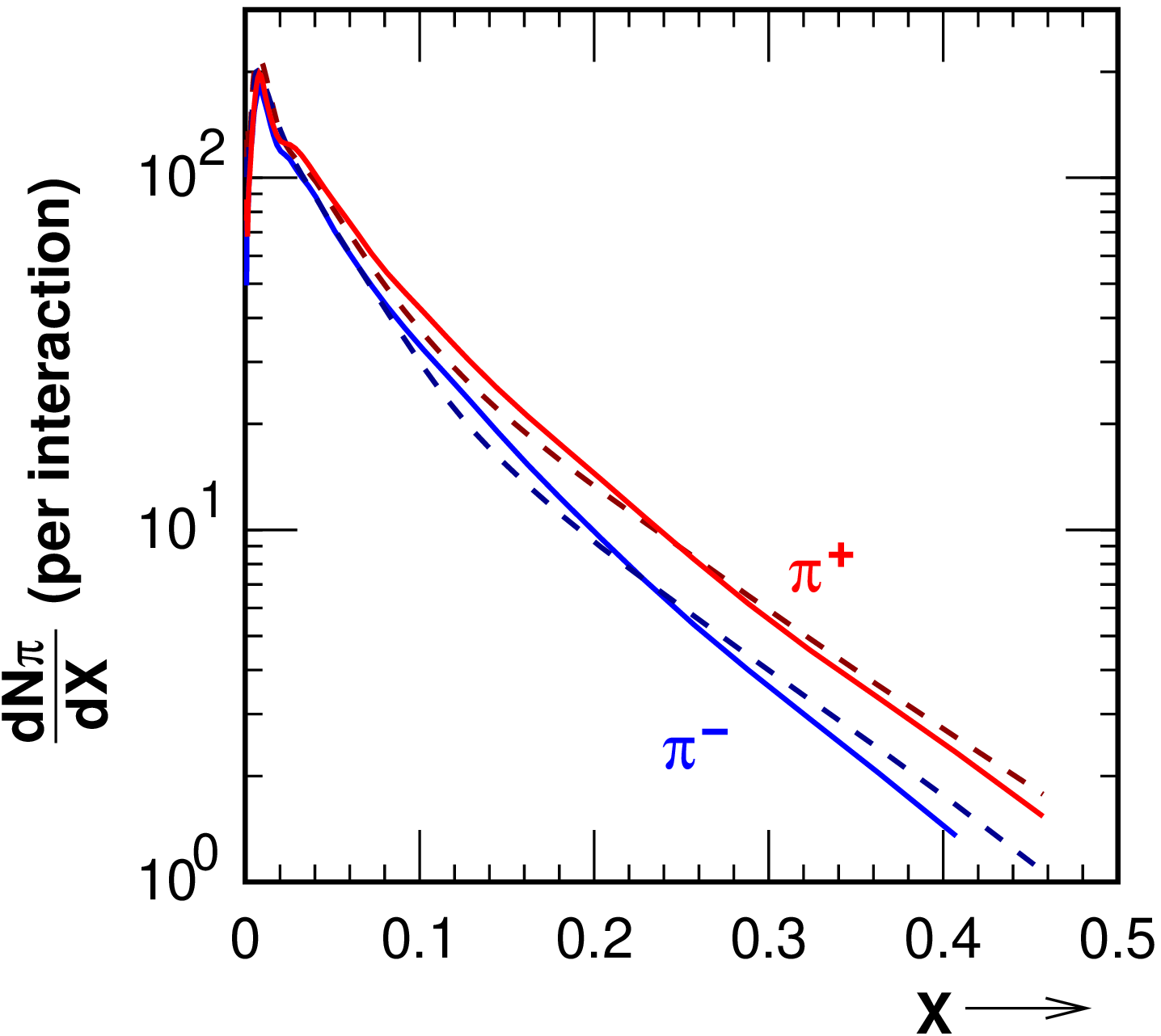}
  }\caption{
Comparison of pion production spectra between 
the original JAM (solid line) and the original DPMJET-III (dashed line)
integrated all over the directions in the left panel,
and those of the modified-JAM (solid line) and the modified-DPMJET-III 
(dashed line) in the right panel at 10 GeV.
Here the $X$ denotes the ratio of secondary
kinetic energy to projectile kinetic energy.
}
\label{fig:pis}
\end{figure}

In Fig.~\ref{fig:pis}, 
we show the comparison of the $x$-spectra ($x\equiv E_{secondary}/E_{projectile}$)
of the 
original JAM and the original DPMJET-III,
and those of the modified-JAM and the modified-DPMJET-III
which we actually used to calculate the fluxes in  
Figs.~\ref{test-tsukuba},~\ref{test-norikura}, and~\ref{test-fort}.
In both panels, we find that JAM shows slightly higher spectra than 
DPMJET-III in $x \lesssim 0.2$.
This is the reason for the higher atmospheric muon flux below 1~GeV$/$c 
with the modified-JAM.
And, our modification method cannot modify DPMJET-III to have this 
high x-spectra
for $x \lesssim 0.2$ and also keep the high spectra for $x \gtrsim 0.2$ 
required by the observed muon flux above 1~GeV.

\section{\label{sec:simulation-scheme} Calculation of Atmospheric Neutrino Flux}

The scheme for calculating the atmospheric neutrino fluxes is almost the 
same as for the previous calculations~\cite{hkkm2004,hkkms2006},
and so we only briefly review it here.
We use the same primary flux model
based on AMS \cite{AMS1p, AMS1He} and  BESS \cite{BESSpHe,BESSTeVpHemu} data,
with a spectral index of $-$2.71 above 100~GeV
(see also Refs.\ \cite{Gaisser-hamburg,Gaisser-Honda}).
We used the 
US-standard '76 model of the atmosphere \cite{us-standard}, 
as the error due to the atmospheric density model 
is sufficiently small for the calculation of atmospheric neutrino 
fluxes~\cite{hkkms2006}.
We use the IGRF2005~\cite{igrf} geo-magnetic model, as the variation due
to changes of geo-magnetic field would be small, 
and only extrapolation
is available to the year 2010 at the time of computation.

For the 3D calculation of the atmospheric neutrino fluxes,
we work in the Cartesian coordinate system which has the origin 
at the center of the Earth, with the $Z$-axis extending to the north pole, and 
we consider the surface of the Earth is a sphere with a radius of 
$R_e=6378.180$~km.
However, the position on the Earth is rather well represented by the 
spherical polar coordinate system $(r, \theta, \phi)$ with $r=R_e$,
related to the Cartesian coordinate system by
\begin{equation}
\begin{array}{c l}
x&=R_e\sin\theta\cos\phi,\\
y&=R_e\sin\theta\sin\phi, \ \ \  {\rm and}\\
z&=R_e\cos\theta\\
\end{array}
\end{equation}
The local coordinate system at the detector is constructed based on this 
polar coordinate system.
The direction of the x-axis is in the increasing direction of $\theta$,
the direction of the y-axis is in the increasing direction of $\phi$, and
the direction of the z-axis is in the increasing direction of $r$.
Therefore, the azimuth angle is measured counterclockwise
from south in the local coordinate system.

In addition to the surface of the Earth, we assume three more spheres; 
the injection sphere, the simulation sphere, and the escape sphere.
In the previous work, we took 
the radius of the injection sphere as $R_{inj}= R_e + 100$~km, 
the radius of simulation sphere as $R_{sim}=R_e + 3000=9378.18$~km, and  
the escape sphere as $R_{esc}= 10 \times R_e=63781.80$~km.
However, we take the radius of the simulation sphere to be   
$R_{sim}=10 \times R_e=63781.80$~km, the same as the escape sphere,
as a result of the increase in computation power.
However, this change does not cause any visible difference in the results.

Cosmic rays are sampled on the injection sphere uniformly 
toward the inward direction, 
following the given primary cosmic ray spectra. 
Before they are fed to the simulation code for the propagation in air, 
they are tested to determine whether they pass the rigidity cutoff, i.e., 
the geo-magnetic barrier.
For a sampled cosmic ray, the `history' is examined by solving the 
equation of motion in the negative time direction.
When the cosmic ray reaches the escape sphere without touching the injection
sphere again in the inverse direction of time, 
the cosmic ray can pass through the magnetic barrier following its trajectory 
in the normal direction of time.

The propagation of cosmic rays is simulated in the space between 
the surface of Earth and the simulation sphere.
When a particle enters the Earth, we discard the particle.
Note, those particles which enter the Earth are mainly muons, and 
the number of muons which enter the Earth are far smaller than the 
number of muons which decay in air.
Also, when muons enter the Earth, they lose energy quickly, and 
decay at rest or are captured by an atom in the rock
and produce a neutrino with energies lower than 100\,MeV.

Note, the neutrino detectors are very small compared with the size 
of the Earth, 
and are considered as infinitesimal points on the 
surface of the Earth.
We need  a finite size ``virtual detector'' instead of the
infinitesimal neutrino detector in the 3D calculation of the
atmospheric neutrino flux.
However, the averaged neutrino flux in the finite size ``virtual detector'' 
is a little different from the neutrino flux at the actual neutrino detector, 
and we introduced the ``virtual detector correction'' in Ref.~\cite{hkkms2006}.
This  ``virtual detector correction'' is refined in this paper.

The procedure for the correction is the same as in Ref.~\cite{hkkms2006}.
Using the inside of the  circle with radius of $\theta_d$ as the
virtual detector,
the flux  $\Phi_{\theta_d}$  obtained with the virtual detector 
and the flux $\Phi_0$ at the real target detector are related by
\begin{equation}
\label{eq:wide_flux}
\Phi_{\theta_d} \simeq \Phi_0 + \Phi_0' \theta_d^2~,
\end{equation}
The term with  $\Phi_0'$  may be cancelled out using two virtual detectors, 
with radii of $\theta_1$ and $\theta_2$, as
\begin{equation}
\label{eq:point_flux}
\Phi_0 \simeq 
\frac
{\theta_1^2\Phi_2  - \theta_2^2\Phi_1}
{\theta_1^2  - \theta_2^2}
=
\frac
{\Phi_2  - r^2\Phi_1}
{1 - r^2},
\end{equation} 
where we assumed $\theta_1 > \theta_2$.

In appendix~\ref{virtual detector}, we study the ratio of the
statistical errors $\Delta\Phi_0/\Delta\Phi_1$ as the function of $r$.
We find that the ratio takes the minimum value 
$\sqrt(5)\simeq 2.236$ at $r=1/\sqrt{2}\simeq 0.707$, 
whereas we took $r=0.5$ in Ref.~\cite{hkkms2006}.
However, the ratio $\Delta\Phi_0/\Delta\Phi_1$ is a slowly varying function 
of $r$ near the minimum, and the ratio at $r=0.5$ is not too bad ($2.517$).

We take the radius of the virtual detector I as 1113.6\,km corresponding to
the angle from the center of the Earth of 10~degrees,
as in Ref.~\cite{hkkms2006},
but we take the radius of the virtual detector II as 787.4\,km corresponding to 
7.071~degrees in this work. 
The ``virtual detector correction'' is applied  
up to $E_\nu=10$~GeV, as
we find a large difference in the fluxes determined by 
the virtual detectors I and II for some directions near the horizon
 (Fig.~\ref{virtual-ratio-ew}, in appendix~\ref{virtual detector}). 

In the previous work, we connected 
the atmospheric neutrino fluxes calculated in the 3D scheme 
to the one calculated in the 1D scheme at 10\,GeV.
In this work, we connected the 3D scheme calculation to 1D one
at 32\,GeV, since we find a dipole-like variation of $\nu_e$ 
and $\bar\nu_\mu$ for horizontal directions due to the muon bending
in the geo-magnetic field at 10\,GeV.

It is not efficient to calculate the atmospheric neutrino spectrum
from 0.1~GeV to 32~GeV in one simulation, 
as the cosmic ray spectra are so steep.
Therefore, we calculate the atmospheric neutrino flux limiting the 
lowest neutrino energy to 1~GeV, 3.16~GeV, 10~GeV, and 31.6~GeV,
also limiting the lowest energy of primary cosmic ray to 1~GeV, 
3.16~GeV, 10~GeV, and 31.6~GeV respectively.
The flux calculated in those simulations are combined into 
single atmospheric neutrino spectrum for each direction 
and each kind of neutrino. 

\begin{figure}[htb]
  \centering{
  \includegraphics[height=2.2in]{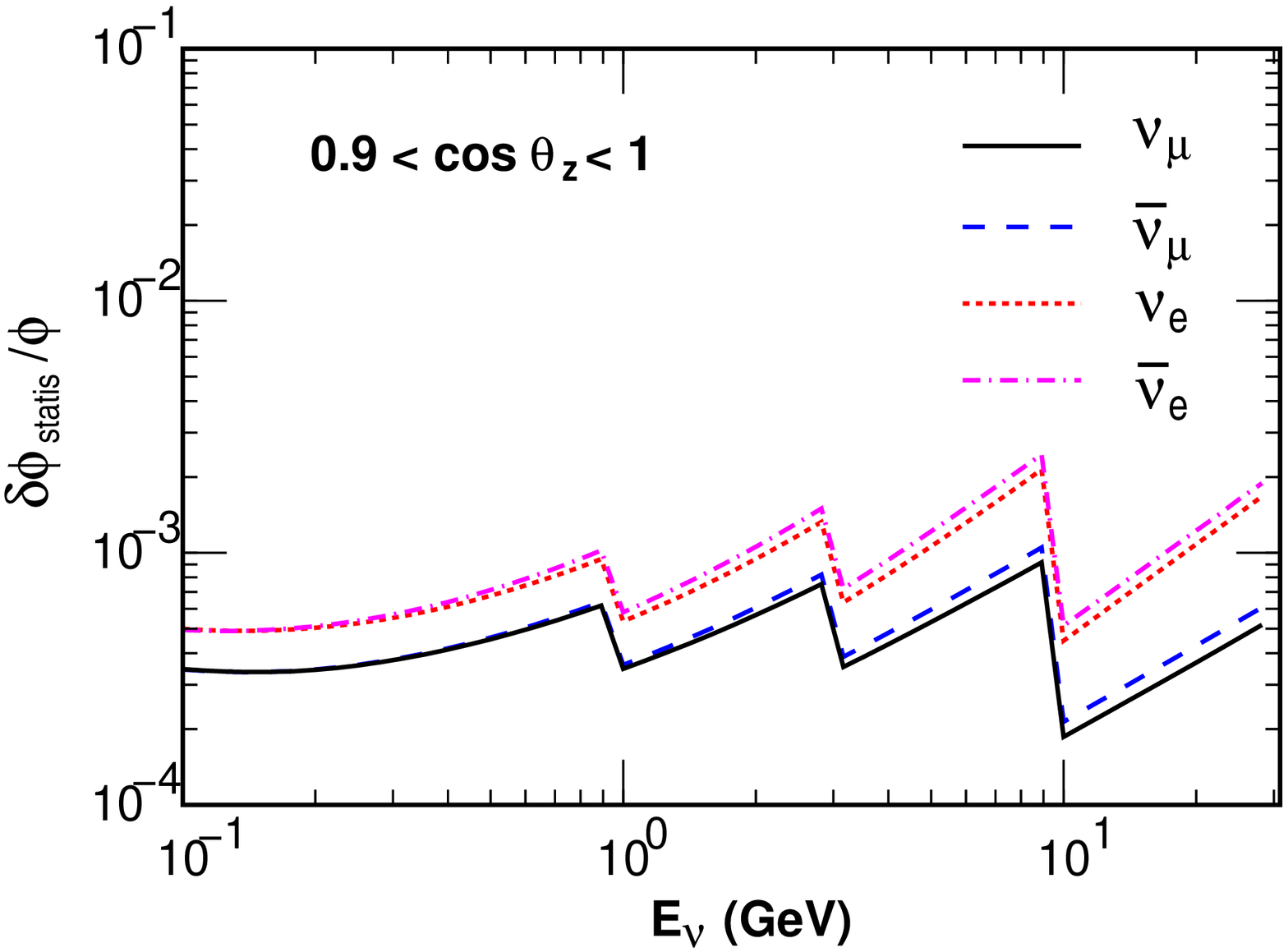}
  \includegraphics[height=2.2in]{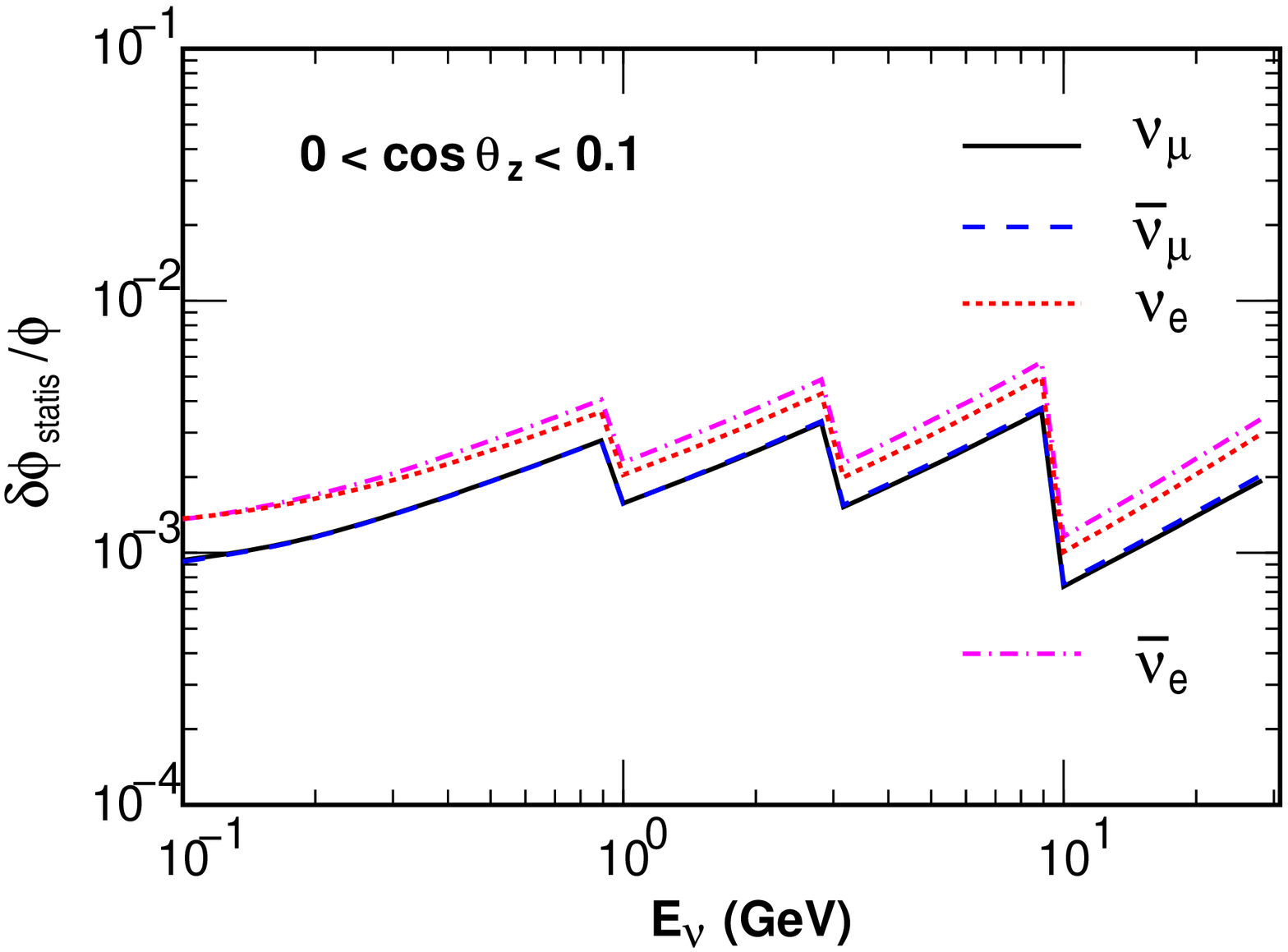}
  }\caption{
The ratio of statistical errors of the atmospheric neutrino flux calculated 
to the flux value in this work.
We show the statistical errors for near-vertical directions 
($0.9 < \cos\theta_z<1$) in the left panel, and for near-horizontal 
directions ($0.9 < \cos\theta_z<1$)  in the right panel, both 
averaging over all azimuthal directions.
}
\label{fig:statistical_error}
\end{figure}

We show the statistical error in this work 
for near-vertical directions and for near-horizontal directions averaging 
over all azimuth angles, in Fig.~\ref{fig:statistical_error}.
For the horizontal directions, the opening angle of the virtual detector
decrease with $\cos\theta$, where $\theta$ is the zenith angle.
Therefore, the statistical error for near-horizontal directions is larger than
that for near-vertical directions.
The statistical error depends not only on the calculation scheme,
but also on the analysis method of the simulation data.
The analysis method of the simulation data and the estimation of 
the statistical error in our calculation is explained in 
appendix~\ref{statistical-error} in some detail.

We find the statistical errors are well below 1~\% for the 
average over the azimuth angles.
For each azimuth bin, however, the statistical errors reach a few percent
for some directions.
The largest statistical errors are found at the energy just below 10~GeV,
due to the virtual detector correction applied up to this energy.

\section{\label{nflx} atmospheric neutrino fluxes calculated with modified-JAM}

\begin{figure}[htb]
  \centering{
  \includegraphics[height=2.5in]{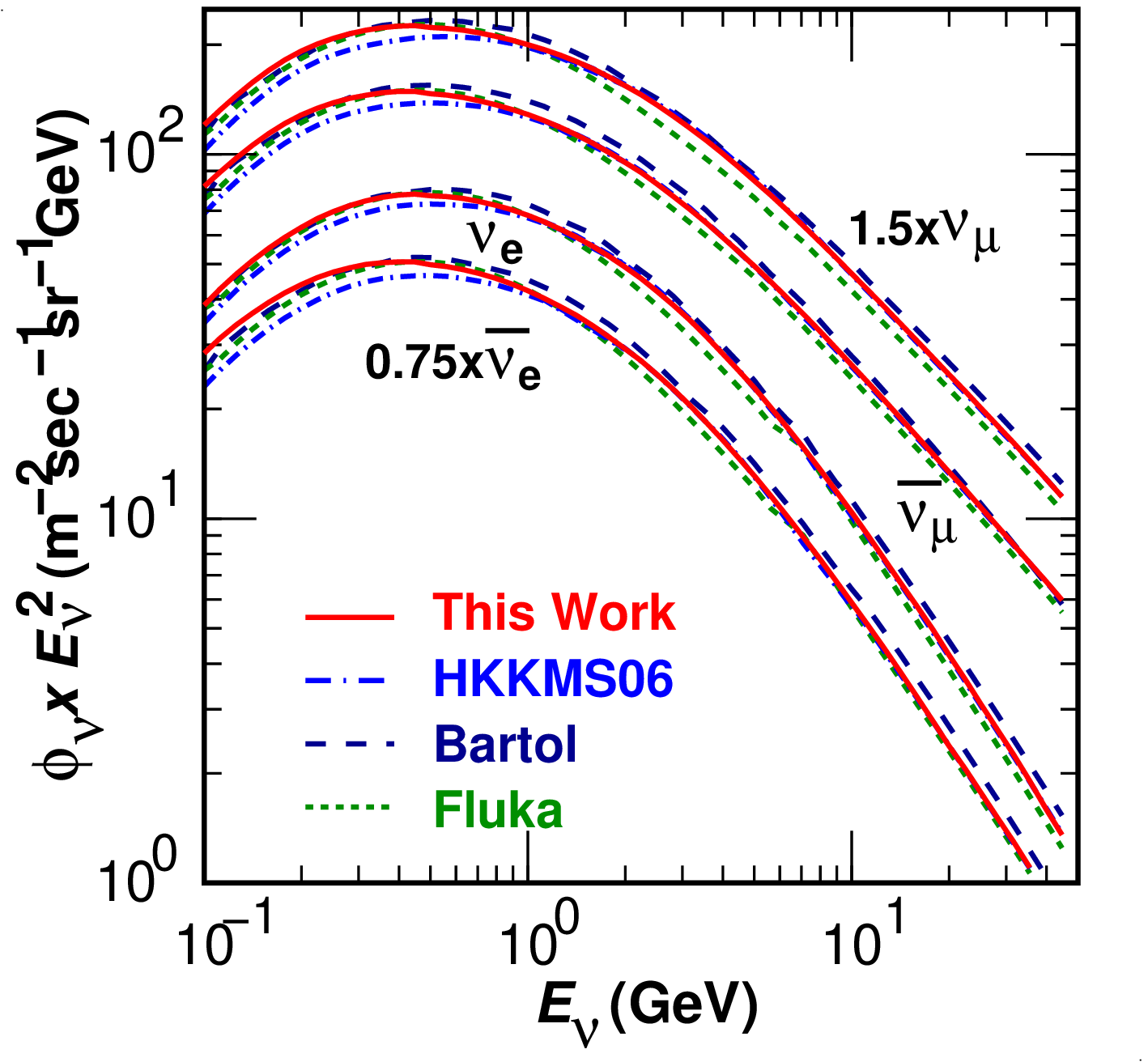}
  \includegraphics[height=2.5in]{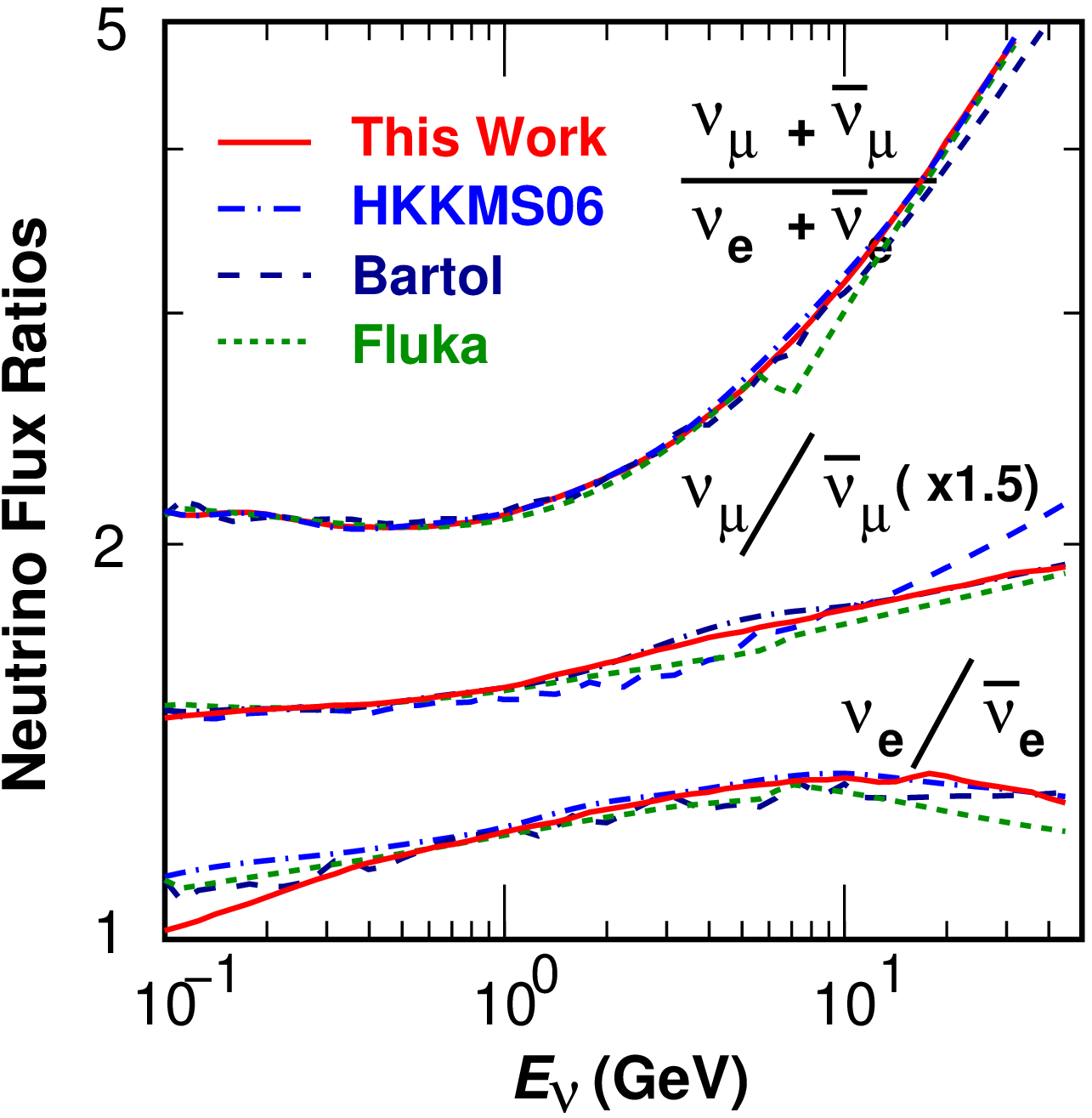}
  }\caption{Comparison of atmospheric neutrino fluxes calculated for Kamioka 
averaged over all 
directions (left panel), and the flux ratios (right panel),
with other calculations.
The dashed lines are the calculation by the 
Bartol group~\cite{barr2004}\cite{barr-web}, 
dotted lines for the FLUKA group~\cite{battis2002}, 
and dash dot for our previous calculation (HKKM06).}
 \label{nflx-kam}
 \end{figure}

\begin{figure}[htb]
  \centering{
  \includegraphics[height=3.5in]{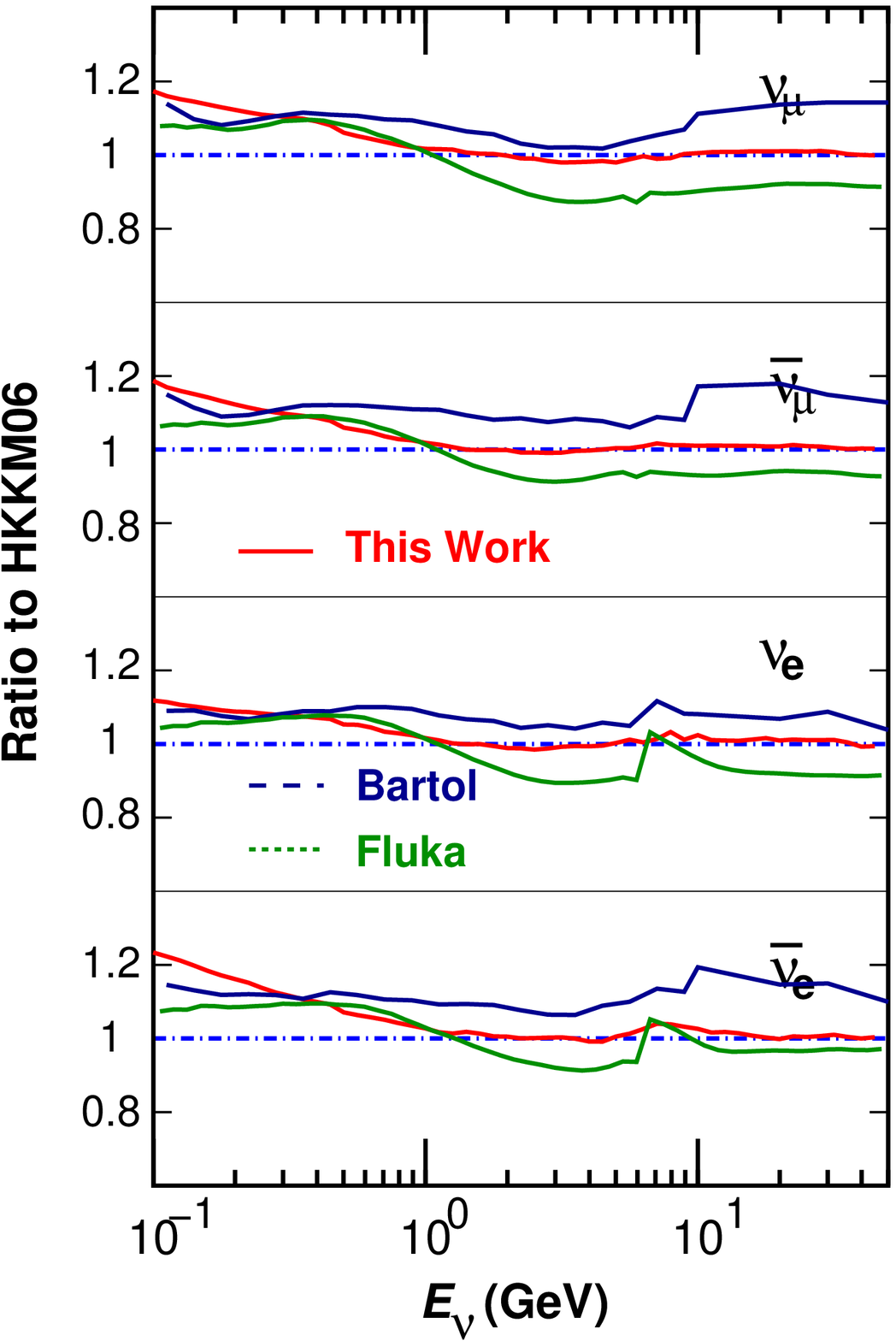}
  }\caption{Atmospheric neutrino fluxes calculated for Kamioka 
averaged over all directions, as a ratio to the HKKM06 calculation.
The dashed lines are the calculation by 
Bartol group~\cite{barr2004}\cite{barr-web} 
and dotted 
lines the FLUKA group~\cite{battis2002}.}
  \label{nflx-ratio2hkkm06}
 \end{figure}

We show the calculated atmospheric neutrino fluxes with the modified-JAM
in this section.
In the left panel of Fig.~\ref{nflx-kam}, the calculated 
atmospheric neutrino spectra are shown averaged over all directions from
0.1~GeV to 32~GeV for Kamioka, together with the ones calculated in our 
previous work, and those 
of the Bartol group~\cite{barr2004}\cite{barr-web},
and the FLUKA group~\cite{battis2002}.
Above 32 GeV, the 3D calculation is smoothly connected to the 1D calculation
carried out in the previous work.
As the modified-JAM is used below 32\,GeV,
any difference above that is due to 
the difference of the calculation scheme between 3D and 1D.
However, the difference between present and previous works is very 
small in the figure above 1\,GeV.
Note, we magnify the difference between present and previous works 
in the ratio in Fig.~\ref{nflx-ratio2hkkm06}.
The difference is less than a few percent above 1~GeV.
On the other hand, 
the atmospheric neutrino fluxes calculated with the modified-JAM show an 
increase from the previous one below 1~GeV,
as is expected from the increase of atmospheric muon spectra below 1~GeV$/$c.

In the right panel of Fig.~\ref{nflx-kam}, we show the ratios of the
atmospheric neutrino flux averaged over all directions.
It is seen that the differences in the flux ratios are very small 
between present and previous works above 1~GeV
as the absolute atmospheric neutrino fluxes.
It is noticeable that the $(\nu_\mu + \bar\nu_\mu)/(\nu_e + \bar\nu_e)$
ratios are very close to each other including the ones calculated 
by different authors.
We find there is a visible difference in $\nu_e/\bar\nu_e$  between
the present and previous works.
As is seen in the left panel of Fig.~\ref{fig:pis} in sec~\ref{intmodel},
the original JAM interaction model has a smaller $\pi^+/\pi^-$ ratio,
and is responsible for the smaller  $\nu_e/\bar\nu_e$ ratio below 1~GeV.
Note, we do not modify the interaction model more than the muon flux data
require..
It is difficult to examine the $\mu^+/\mu^-$ ratio at the momenta
corresponding to the $\nu_e/\bar\nu_e$  below 1~GeV,
due to the small statistics in the observation at balloon altitude.

The muons at sea level or mountain altitude are not useful to examine
the atmospheric neutrino of these energies, since the muons
result from
higher energy pions at higher altitude.

 \begin{figure}[htb]
  \centering{
  \includegraphics[height=2.7in]{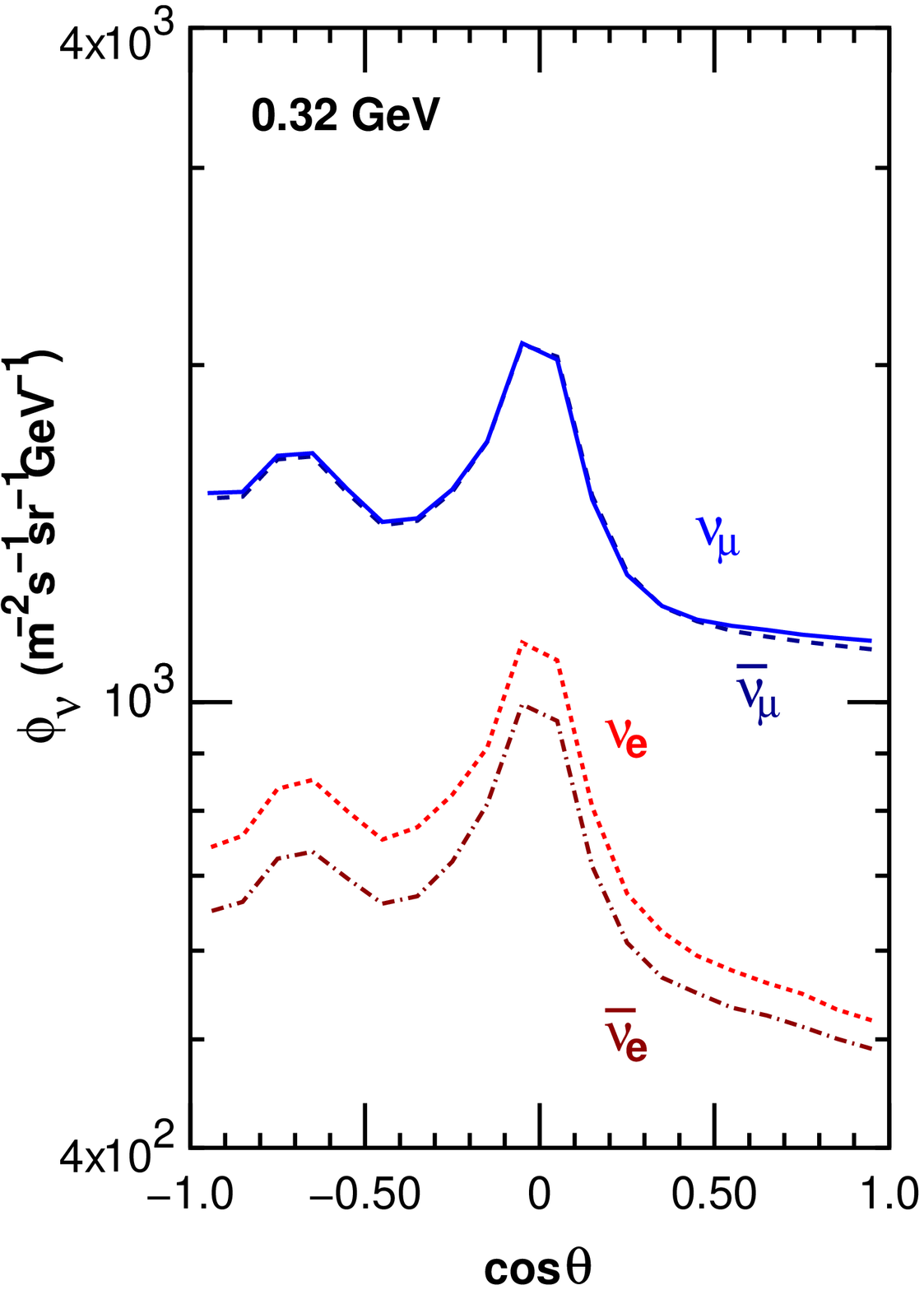}\hspace{5mm}
  \includegraphics[height=2.7in]{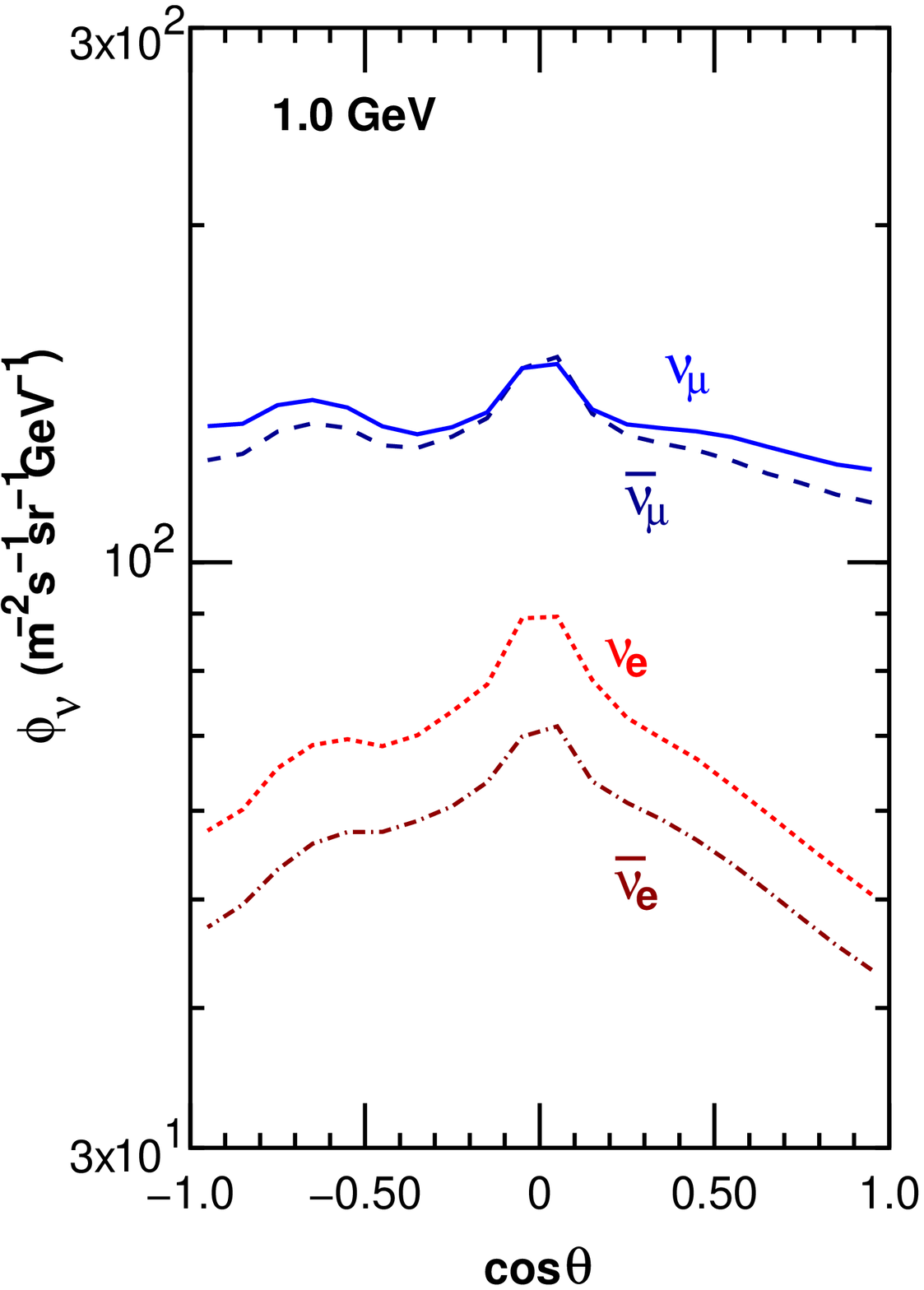}\hspace{5mm}
  \includegraphics[height=2.7in]{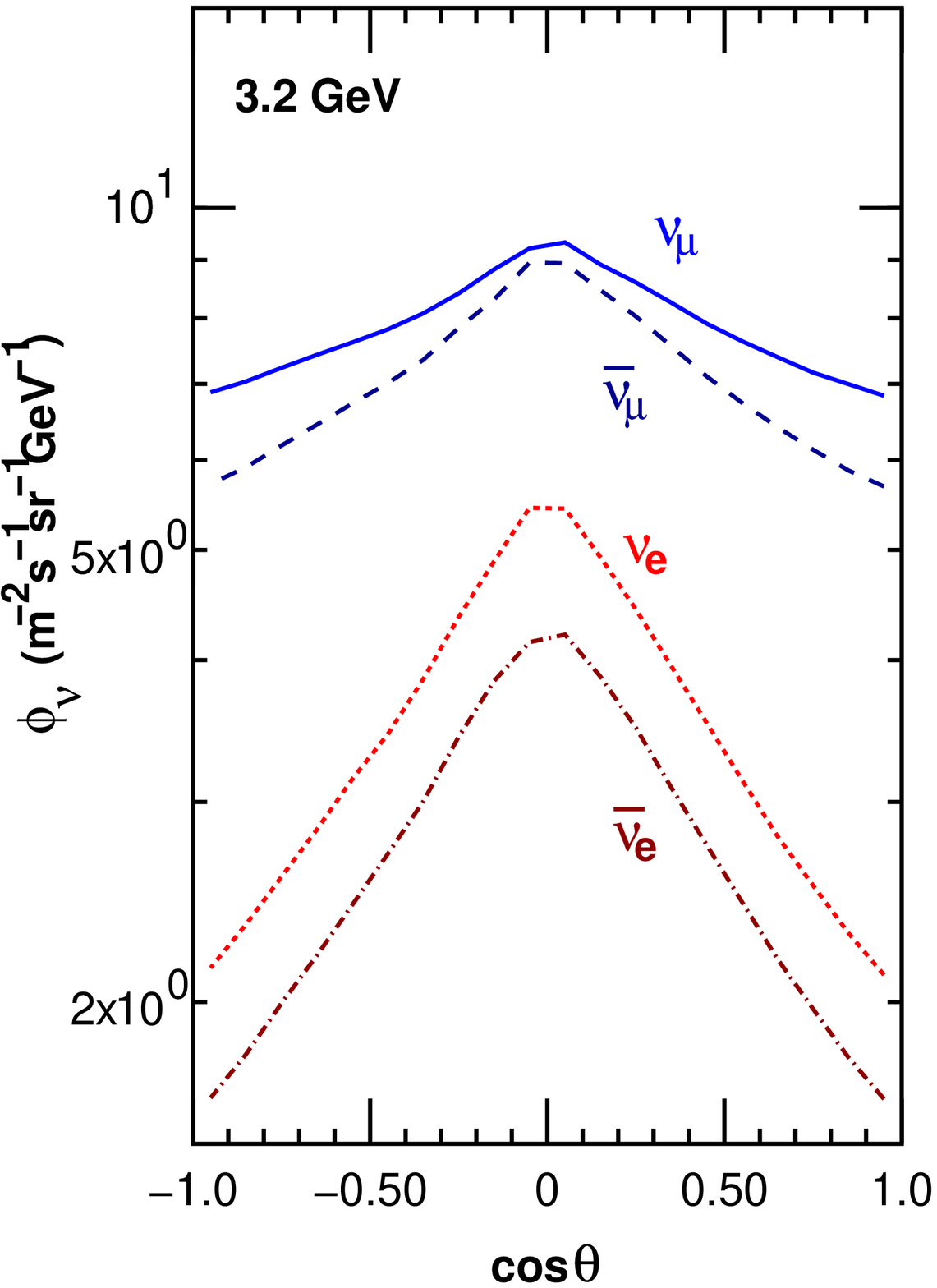}
  }\caption{
The zenith angle dependence of atmospheric neutrino flux averaged 
over all azimuthal angles calculated for Kamioka.
Here $\theta$ is the arrival direction of the neutrino,
with $\cos\theta=1$ for vertically downward going neutrinos,
and  $\cos\theta=-1$ for vertically upward going neutrinos.
}
  \label{kam-zdep}
 \end{figure}

 \begin{figure}[htb]
  \centering{
  \includegraphics[height=2.7in]{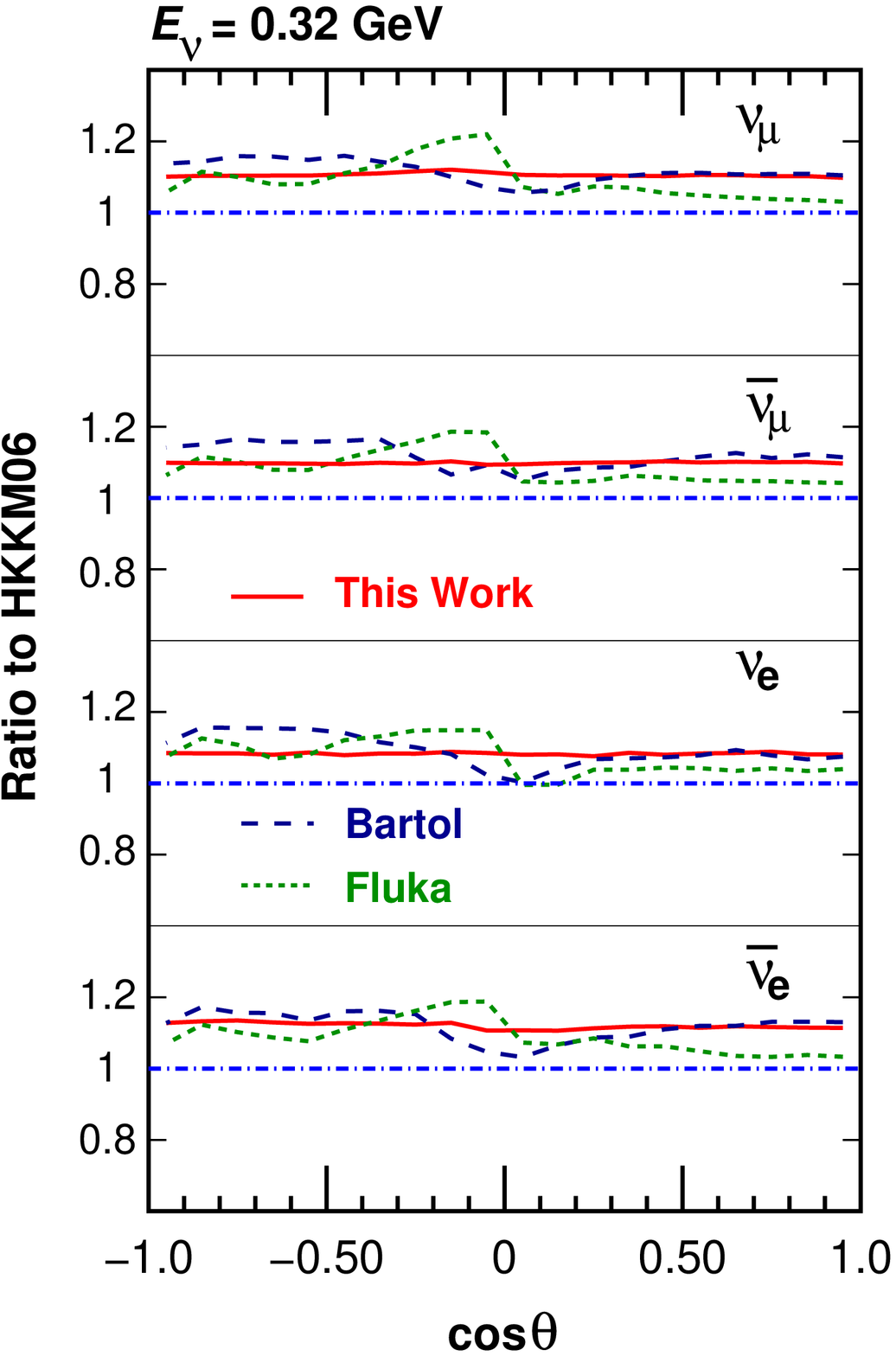}
  \includegraphics[height=2.7in]{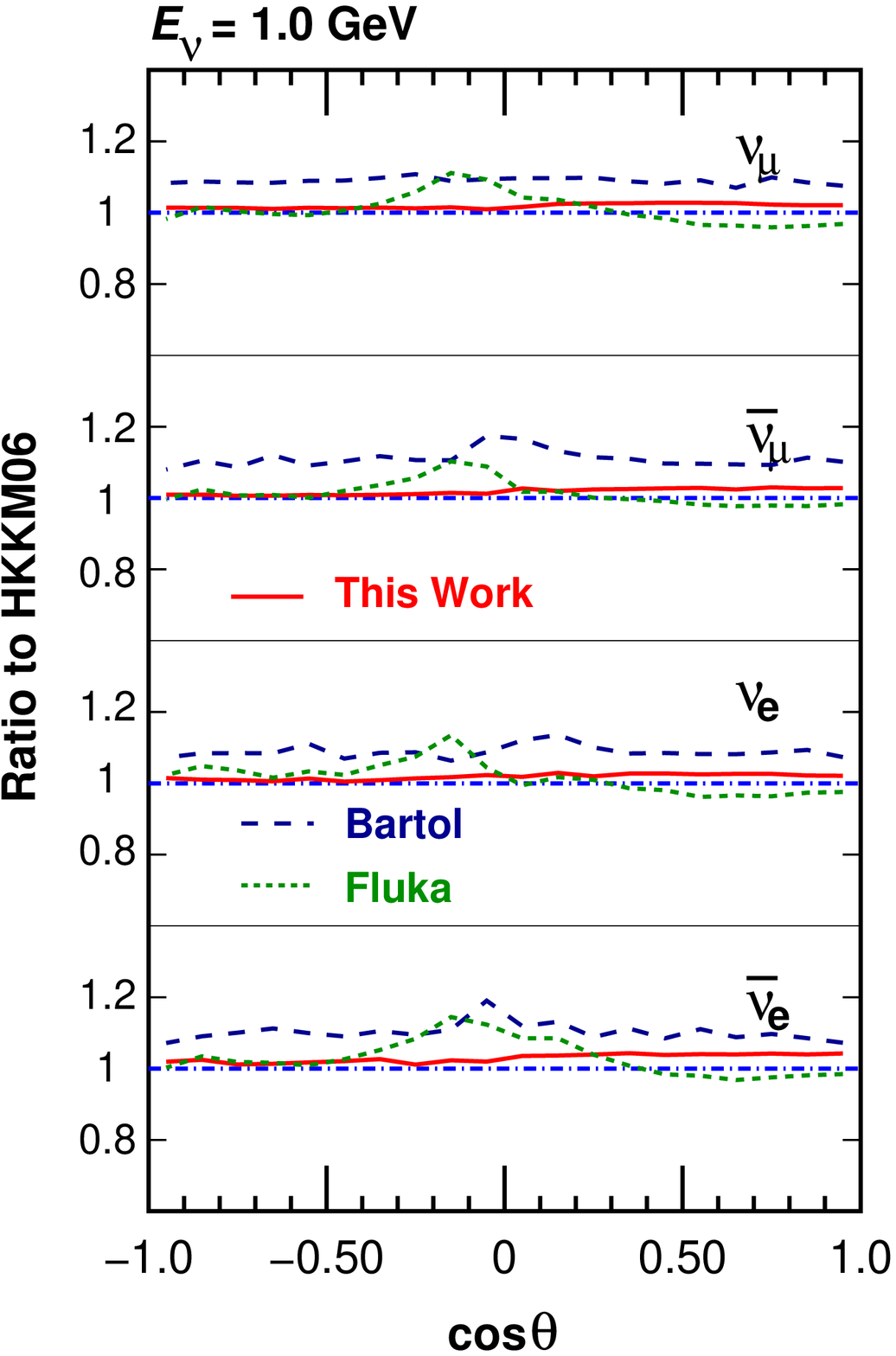}
  \includegraphics[height=2.7in]{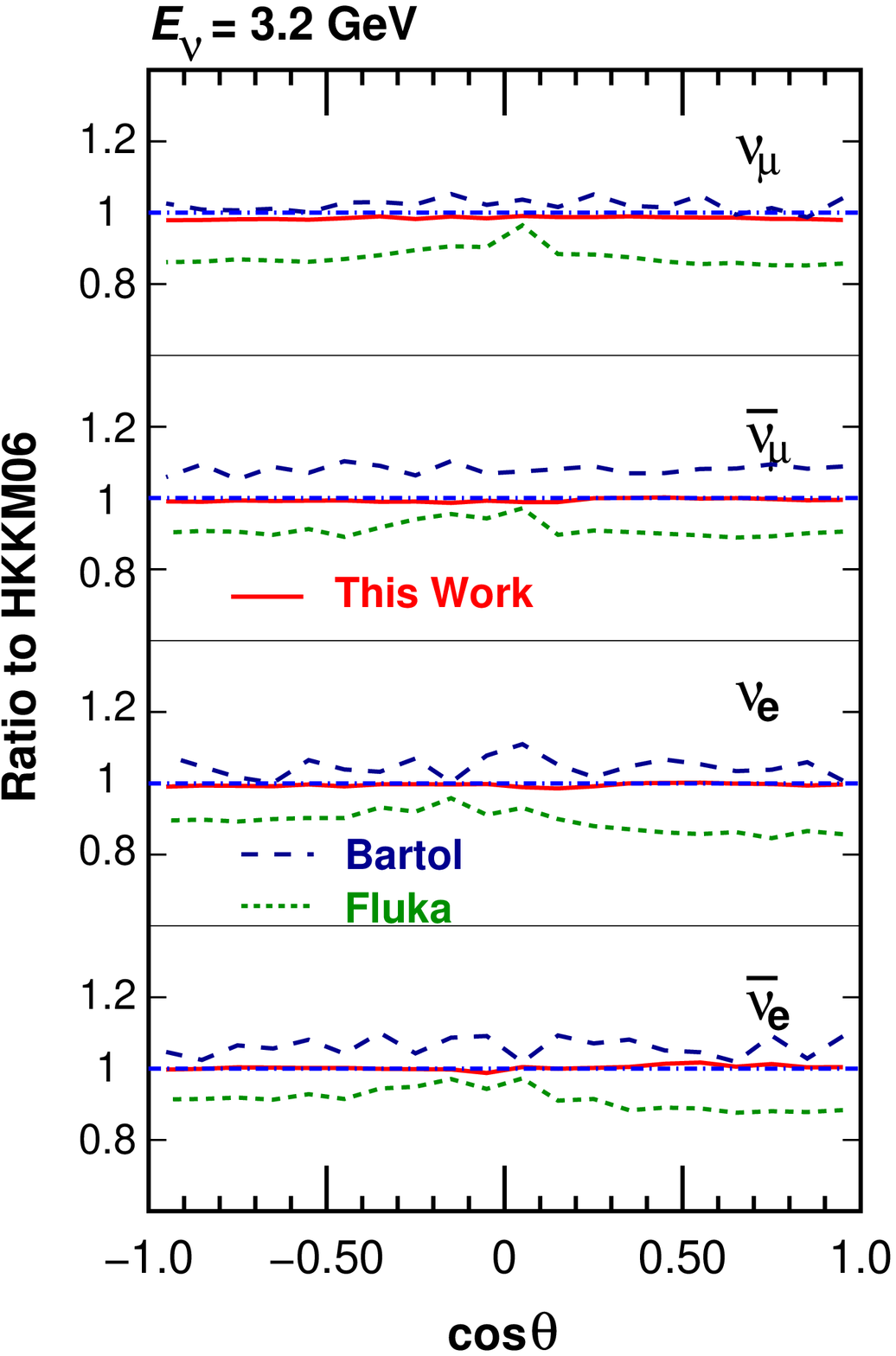}
  }\caption{
Ratio of the atmospheric neutrino flux averaged over azimuth angles
as the function of zenith angle to the same quantity calculated in HKKM06.
The short dash lines stand for the calculation by 
Bartol group~\cite{barr2004}\cite{barr-web}  and dotted 
lines for the FLUKA group~\cite{battis2002}.
}
  \label{kam-zdep-ratio}
 \end{figure}

In Fig.~\ref{kam-zdep} we show the atmospheric neutrino fluxes as a function
of the zenith angle averaging over all the azimuthal angles 
at 3 neutrino energies; 0.32, 1.0, and 3.2\,GeV for Kamioka.
In Fig~\ref{kam-zdep-ratio} we show the comparison of the present and 
previous works in the ratio as the function of zenith angle.
There is a difference due to the increase of the neutrino flux itself,
but the ratio is almost constant.
Actually, the calculated zenith angle dependences are virtually the 
same as for the calculation in Ref~\cite{hkkm2004}.

 \begin{figure}[htb]
  \centering{
  \includegraphics[height=3.0in]{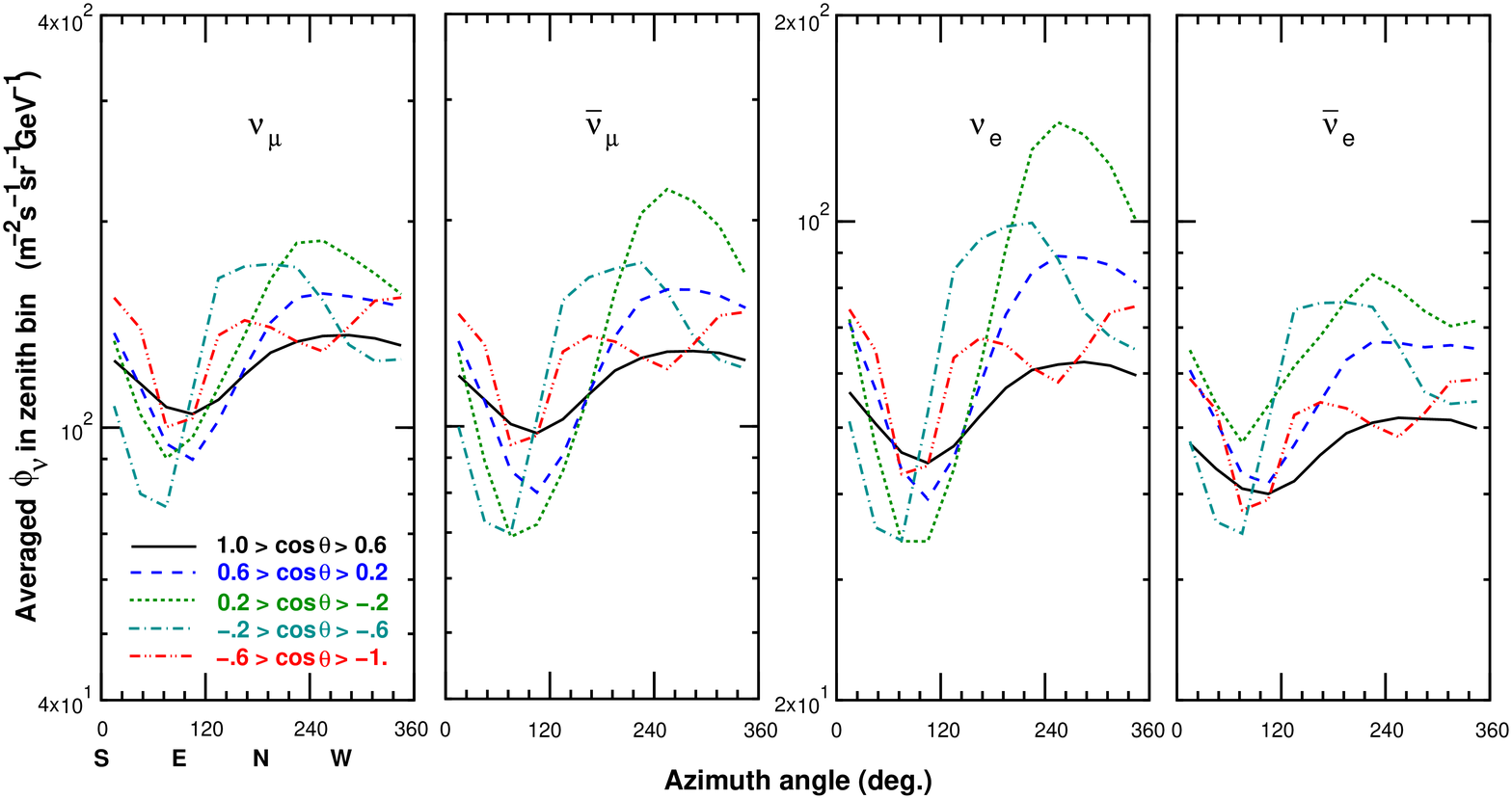}
  }\caption{
The azimuthal angle dependence of atmospheric neutrino flux at 1 GeV,
averaged 
over zenith angle bins of $1>\cos\theta>0.6$, $0.6>\cos\theta>0.2$, 
$0.2>\cos\theta>-0.2$, $-0.2>\cos\theta>-0.6$, and $-0.6>\cos\theta>-1$
calculated for Kamioka.
}
  \label{kam-adep10}
 \end{figure}

It seems that the zenith angle dependence of the 3D calculation smoothly 
connected to that of the 1D calculation just above 3.2\,GeV
for the average over all azimuth angles. 
However, this is not true when we study the variation of atmospheric 
neutrino flux as a function of azimuthal angle.
In Fig.~\ref{kam-adep10} we show the variation of atmospheric neutrino flux 
as the function of the azimuthal angle averaging them over the five zenith angle ranges,
$1>\cos\theta>0.6$, $0.6>\cos\theta>0.2$, $0.2>\cos\theta>-0.2$, 
$-0.2>\cos\theta>-0.6$, and $-0.6>\cos\theta>-1$, at 1~GeV for Kamioka.
It is seen in that the variation of the atmospheric neutrino flux has  
complex structures at 1\,GeV due to the
rigidity cutoff and  muon bending in the geo-magnetic 
field~\cite{Lipari:2000du}.
Note, the variation of upward going neutrinos
($-0.2 > \cos\theta$) 
is much more complicated than the variation of downward going neutrinos
($\cos\theta > 0.2$).
This is because the upward going neutrinos 
are produced in a far larger area on the Earth than the 
downward going neutrinos, and are affected by large variation of rigidity cutoff
and geo-magnetic field.
The downward going neutrinos are produced just above the detector.

 \begin{figure}[htb]
  \centering{
  \includegraphics[height=3.0in]{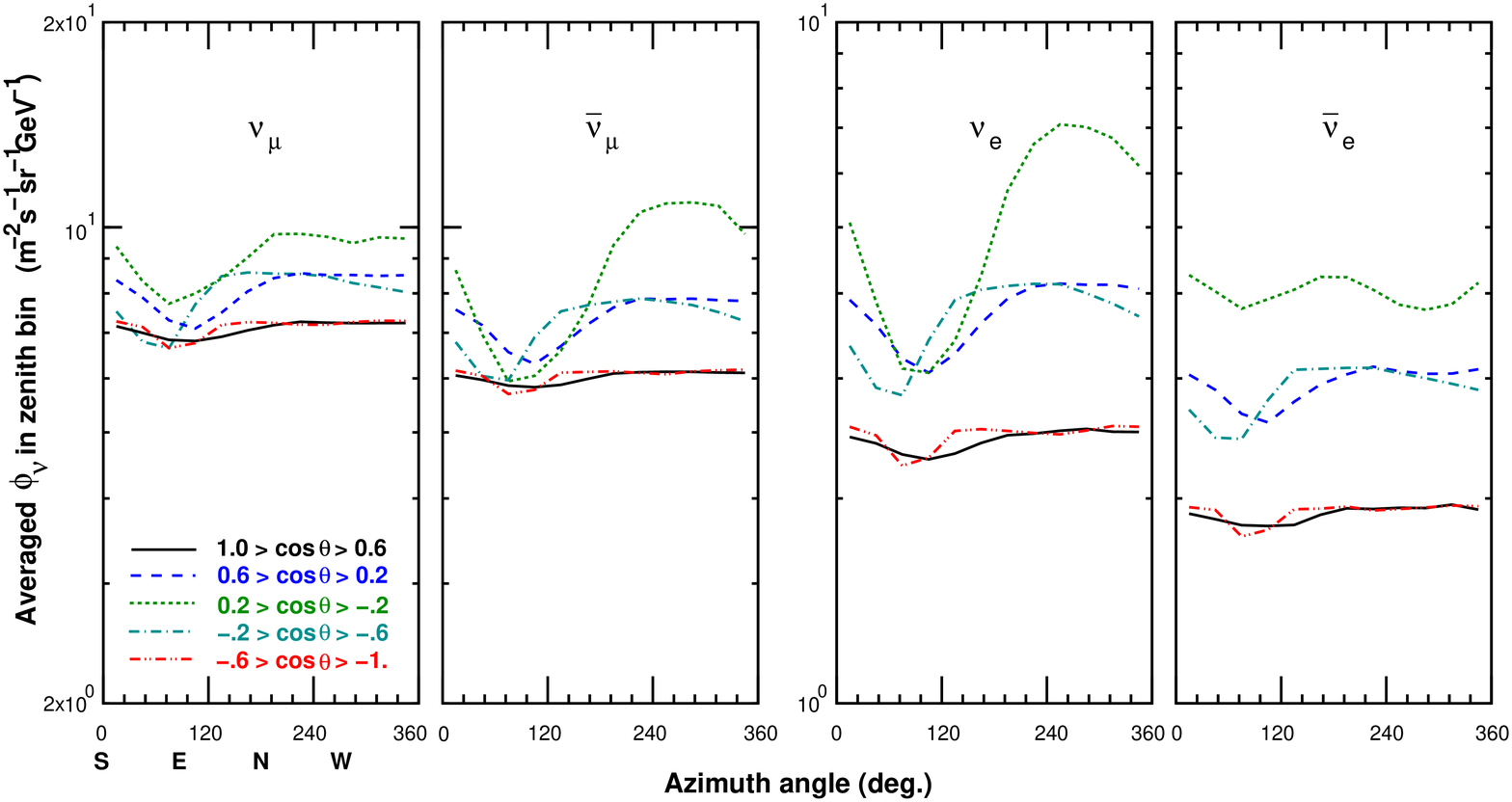}
  }\caption{
The azimuthal angle dependence of atmospheric neutrino flux at 3.2\,GeV,
averaged 
over zenith angle bins of $1>\cos\theta>0.6$, $0.6>\cos\theta>0.2$, 
$0.2>\cos\theta>-0.2$, $-0.2>\cos\theta>-0.6$, and $-0.6>\cos\theta>-1$
calculated for Kamioka.
}
  \label{kam-adep15}
 \end{figure}

In Fig.~\ref{kam-adep15} we show the variation of atmospheric neutrino flux as 
the function of the azimuthal angle averaging them over the same five zenith angle bins
at 3.2~GeV for Kamioka.
At this neutrino energy, we find the deficit of neutrino flux due to the 
rigidity cutoff from the East direction ($\phi \sim 90^\circ$) clearly for
all kind of neutrinos and all zenith angle bins.
However, the $\nu_e$ and $\bar\nu_\mu$ show large dipole structures in the 
horizontal zenith angle bin  ($0.2>\cos\theta>-0.2$) due to 
the muon bending in the geo-magnetic field.

 \begin{figure}[htb]
  \centering{
  \includegraphics[height=3.0in]{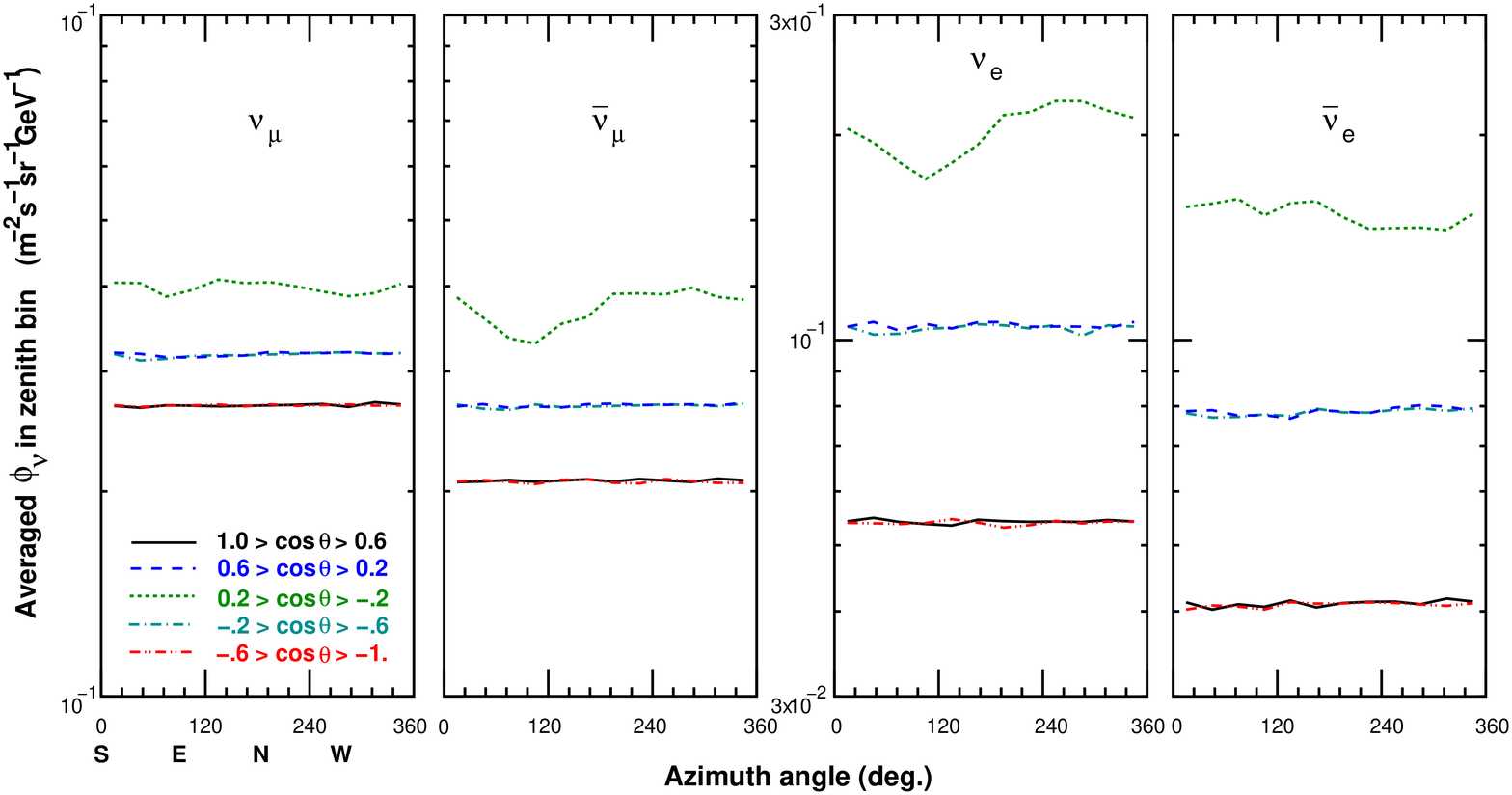}
  }\caption{
The azimuthal angle dependence of atmospheric neutrino flux at 10\,GeV,
averaged 
over zenith angle bins of $1>\cos\theta>0.6$, $0.6>\cos\theta>0.2$, 
$0.2>\cos\theta>-0.2$, $-0.2>\cos\theta>-0.6$ and $-0.6>\cos\theta>-1$
calculated for Kamioka.
}
 \label{kam-adep20}
 \end{figure}

In Fig.~\ref{kam-adep20} we show the variation of atmospheric neutrino flux as 
a function of the azimuthal angle averaging over the same five zenith angle bins
at 10~GeV for Kamioka.
At this energy, the atmospheric neutrinos seem to be free from the rigidity
cutoff, but 
the $\nu_e$ and $\bar\nu_\mu$ still show the dipole structures in the 
horizontal zenith angle bin  ($0.2>\cos\theta>-0.2$).

 \begin{figure}[htb]
  \centering{
  \includegraphics[height=3.0in]{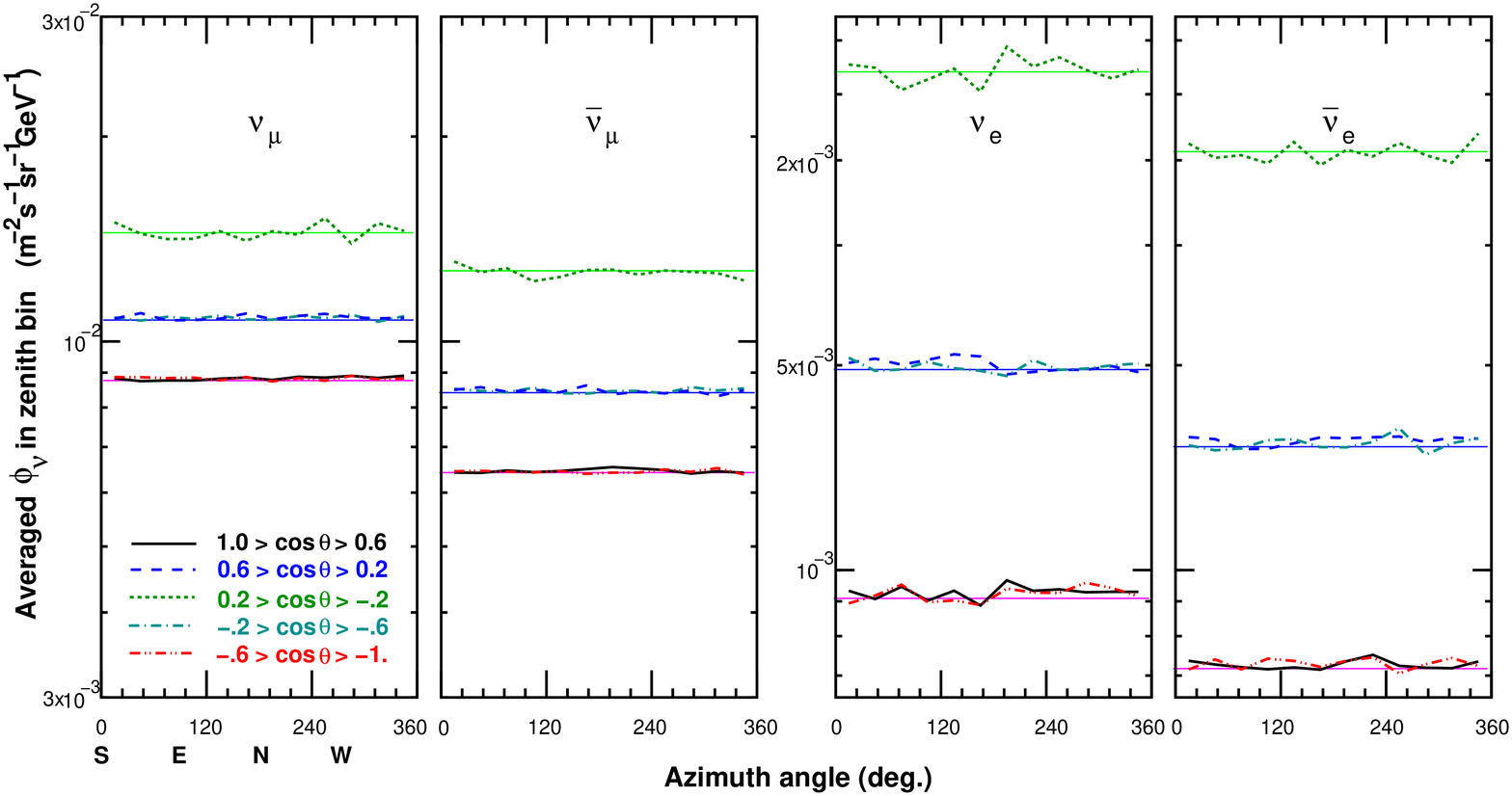}
  }\caption{
The azimuthal angle dependence of atmospheric neutrino flux at 32 GeV,
averaged 
over zenith angle bins of $1>\cos\theta>0.6$, $0.6>\cos\theta>0.2$, 
$0.2>\cos\theta>-0.2$, $-0.2>\cos\theta>-0.6$, and $-0.6>\cos\theta>-1$ 
calculated for Kamioka.
The straight lines in the figure show the flux values calculated in 
the 1D scheme.
}
  \label{kam-adep25}
 \end{figure}

The dipole structure of $\nu_e$ and $\bar\nu_\mu$ 
seems to extend to a few 10~GeV in the horizontal bin.
In Fig.~\ref{kam-adep25}, we show the variation of atmospheric neutrino flux as 
the function of the azimuthal angle averaging them over the 
same five zenith angle bins
at 32~GeV
calculated for Kamioka.
We still see the dipole azimuth angle variation of $\nu_e$ at this energy
in the horizontal bin.
However, the amplitude of the variation is 
almost the same as that of the statistical error.
Therefore, we connect the atmospheric neutrino flux calculated in 
3D scheme to the ones calculated in 1D scheme at 32 GeV.
In Fig.~\ref{kam-adep25}, we also plotted the flux vales calculated in the
1D scheme. 
We can say the flux values calculated in the 3D and 1D schemes agree
within the statistical error, except for the dipole structure of 
$\nu_e$ in the horizontal bin.

\section{summary and discussion}

After the previous publication, we have searched for an interaction model 
which has better behavior at low energies than DPMJET-III and NUCRIN, 
and we found that the JAM interaction model, used
in PHITS (Particle and Heavy-Ion Transport code System)~\cite{phits}.
It shows slightly better agreement with the HARP experiment than DPMJET-III.

Constructing the inclusive interaction model from the JAM interaction model,
we have used it as the hadronic interaction code in the simulation of cosmic ray 
propagation through the atmosphere below 32\,GeV.
Above this energy, we continue to use the modified DPMJET-III,
since we could reproduce the result of atmospheric muon observations 
by BESS for $\gtrsim$~1~GeV$/$c in the previous work.

Applying some modifications to the inclusive JAM, we were able to increase
the atmospheric muon flux below 1\,GeV$/$c, 
and achieved better agreement with  
BESS observations at Tsukuba (30\,m a.s.l.) and at Fort Sumner (balloon 
altitudes).
However, the increase of the atmospheric muon flux below 1\,GeV$/$c 
made the difference larger between calculation and observations
at Mt.\ Norikura (2770\,m a.s.l.).
We consider the agreement at the balloon altitudes to be more important, since
muons are observed there with almost the same momentum as they are 
produced.
We then calculate the atmospheric neutrino fluxes with the modified-JAM as
the hadronic interaction code below 32\,GeV.

The calculation scheme is essentially the same as in previous calculations,
but the radius of the simulation sphere has been made the same as that 
of the escape sphere, to reduce the boundary effect.
However, there is no visible effect from this change 
in the calculated atmospheric neutrino fluxes.
Also we revised the formula for the size correction of the virtual detector 
to minimize the statistical errors, which were reduced by $\sim$11\%.

Some features of the calculated atmospheric neutrino flux are 
presented in section~\ref{nflx}.
The use of the modified-JAM increased the atmospheric neutrino fluxes 
below 1~GeV corresponding to the increase of atmospheric muon fluxes 
below 1~GeV$/$c.
However, the ratios of the atmospheric neutrinos,
especially  $(\nu_\mu + \bar\nu_\mu)/(\nu_e + \bar\nu_e)$, are 
very close to the previous calculation.
Only the ratio $\nu_e/\bar\nu_e$ shows a decrease near 0.1~GeV due to the
small $\pi^+/\pi^-$ ratio of the original JAM.
It is difficult to study the  $\pi^+/\pi^-$ ratio from the observed 
atmospheric muons.
Also the zenith angle variations of the atmospheric neutrinos are 
virtually the same as the previous calculations.

The azimuth angle variations of the atmospheric neutrino fluxes are studied in 
detail in this paper.
For horizontal directions ($0.2>\cos\theta>-0.2$),
$\bar\nu_\mu$ and $\nu_e$ still show a dipole-like variation even at 10~GeV
due to the deflections of muons in the geo-magnetic field.
However, the magnitude of the variation become almost the same as the 
statistical error at 32~GeV even for horizontal directions.
As a result, we connected the atmospheric neutrino fluxes calculated in the
3D calculation scheme to those calculated in the 
1D calculation scheme at 32\,GeV.
They agree with each other within the statistical error at this energy.

The study of the interaction model and atmospheric muons,
similar to that in the previous publication,
suggests that the magnitude of the error of the atmospheric neutrino flux 
due to the hadronic interaction model at energies below 1\,GeV is 
similar to the maximum difference between the calculated atmospheric muon 
fluxes and observed ones below $\lesssim$ 3~GeV$/$c at balloon altitudes.
Assuming $\sim$10\% of the uncertainty for the total interaction cross section 
of cosmic rays and air nuclei, or the interaction mean free path of the cosmic
rays, we estimate the
uncertainty of the atmospheric neutrino flux below 1~GeV is around 15\%
at 0.3~GeV, and 20\% at 0.1~GeV. 

\begin{acknowledgments}
We greatly appreciate the contributions of J.~Nishimura and A.~Okada
to this paper.
We are grateful to T.~Sanuki, K.~Abe, and
P.G.~Edwards for discussions and comments. 
We also thank the ICRR of the University of Tokyo, 
especially for the use of the computer system.
\end{acknowledgments}

\bibliography{nflx-jam}

\appendix
\section{\label{inclusive-code}Interaction model and calculation of atmospheric neutrino flux}

Here, we explain the construction of the inclusive code from the output of
the original interaction code.
First we ran the original code many times, 
and recorded the kind, energy, and momentum of the secondary particle.
We sort and number them in the ascending order of 
$x \equiv E_{secondary}/E_{projectile}$  as
\begin{equation}
\label{sorted-x}
x_1 < \cdots < x_{i-1} < x_i < x_{i+1} < \cdots < x_n\ \ ,
\end{equation}
for each kind of secondary particle.
Considering this series as a function of integer in $[1,n]$,
we can define a function of a discrete variable in $(0,1]$ as
\begin{equation}
\label{fit-fun1}
x_i = X(\frac{i}{n})\ \ .
\end{equation}
This $X(i/n)$ converges to a continuous function in $[0,1]$ as 
the number of trials, or $n$, increases, and
the asymptotic continuous function can be approximated by a 
proper fitting function for sufficiently large $n$.

Writing the asymptotic function also as  $X(u)$,
we can generate the $x$ following the $x$-distribution of the 
original interaction model with uniform random numbers $u$ in $[0,1]$ as
\begin{equation}
\label{regeneration}
x =X(u)\ \ .
\end{equation}
It is straightforward to calculate the energies of the secondary particles
from ${x}$.

As the fitting function, we used the B-spline function for DPMJET-III
and NUCRIN, but
we take poly-quadra function for JAM for accuracy, 
although the difference in the actual usage is small.

We fit the angular distribution by the fitting formula,
\begin{equation}
\label{a-fit}
F(\cos\theta)d\cos\theta = 
\exp\left(A+ a(\cos\theta+1)^\alpha (1-\cos^2\theta)^\beta  \right)
d\cos\theta \ \ ,
\end{equation}
at low energies ($\lesssim$~32~GeV).
We determine the parameters, $A,a,\alpha$, and $\beta$ for each $x$-bin
with $\log\Delta x = 0.1$.
For higher energies, we fit the angular distribution with a simple 
formula
\begin{equation}
\label{a-fit-simple}
F'(\cos\theta)d\cos\theta =  \exp(A+ a \cdot \cos\theta)d\cos\theta ~,
\end{equation}
which is a special case of eq.~\ref{a-fit}.

Repeating the above procedures for all kinds of secondary particle 
and for all the projectile energy grids,
the inclusive code is constructed for the interaction model.
Note,
the projectile energy grids are selected to cover the required projectile 
energy range.
For the projectile energy in between two energy grids,
we use the set of parameters at either grid energy.
The grid energy to use is determined by the distances to 
both grid energies and a random number in a probabilistic way.

As the grid projectile energies, we took
$\log(E_g/1~{\rm GeV})=$ 0.5, 1.0 ... 5.5, and 6.0 for DPMJET-III, 
and $\log(E_g/1~{\rm GeV})=$ -0.7, -0.6, -0.4, .0, and 0.5 for NUCRIN. 
Fro JAM interaction model, we take
$\log (E_g/1~{\rm GeV}) =$ -1, -0.9, ... 0., 0.1, ...1.6, and 1.7.

As noted previously, 
the inclusive code allows us to readily modify the
secondary spectra of the original interaction model.
When we apply a modification to the inclusive code,
we use replace the uniform random generator to the weighted random number
generator in $[0,1]$,
instead of the uniform random number generator.
We take a simple form for the weight function,
\begin{equation}
w_i(u) = 1 + a_i \cdot u~~,
\end{equation}
and request the random numbers are generated in the probability
proportional to $w(u)$.
The ${a_i}$ are the modification parameter defined at the projectile energy
grid for all the kind of secondary particles.
At each projectile energies, the ${a_i}$ are not independent but the
total energy of secondary particle is conserved on average.
The multiplicities are not modified.

The above modification procedure is an effective way to modify so 
called $Z$-factor ($Z \equiv <x^{1.7}>$, with 
$x \equiv E_{secondary}/E_{projectile}$)
which is a good description for atmospheric muons and neutrinos at high energies
\cite{Gaisser-semi-analytic}.
%

\section{\label{virtual detector} ``Virtual detector correction'' and the optimization}

We have shown in Ref.~\cite{hkkms2006} that the average flux 
($\Phi_{\theta_d}$) 
in a virtual detector with radius of $\theta_d$ is expressed as
\begin{equation}
\label{eq:wide_fluxA}
\Phi_{\theta_d} \simeq \Phi_0 + \Phi_0' \theta_d^2~,
\end{equation}
where $\Phi_0$ is the atmospheric neutrino flux at the target detector,
and $\Phi_0'$ is a constant determined for the target detector.

As in Ref.~\cite{hkkms2006}, we may cancel out the term with  $\Phi_0'$ 
using two averaged fluxes in two concentric virtual 
detectors around the target detector.
Assuming two virtual detectors with radii $\theta_1$ and $\theta_2$,
we can calculate the flux at the target detector ($\Phi_0$) as
\begin{equation}
\label{eq:point_fluxA}
\Phi_0 \simeq 
\frac
{\theta_1^2\Phi_2  - \theta_2^2\Phi_1}
{\theta_1^2  - \theta_2^2}
=
\frac
{\Phi_2  - r^2\Phi_1}
{1 - r^2},
\end{equation} 
where, 
$\Phi_1$ is the averaged flux in the virtual detector I with 
a radius of $\theta_1$, and 
$\Phi_2$ is the averaged flux in the virtual detector II with 
a radius of $\theta_2$.
Here, we assumed $\theta_1 > \theta_2$

The treatment of the virtual detector in Ref.~\cite{hkkms2006} 
corresponds to taking $r=0.5$ in \ref{eq:point_flux}.
However, this value is not optimal for the statistical error due to  
fluctuations in the number of neutrinos observed in the virtual detector.
The fluxes $\Phi_1$ and $\Phi_2$ are calculated as 
$$
\Phi_1 = \frac{N_1}{T \pi \theta_1^2} ,\ \ \ \ {\rm and}\ \ \ \ 
\Phi_2 = \frac{N_2}{T \pi \theta_2^2},
$$
where $T$ is the corresponding observation time of the simulation, and
$N_1$ and $N_2$ are the observed numbers of neutrinos in the virtual
detectors I and II respectively.
The ``true'' flux $\Phi_0$ is calculated from $N_1$ and $N_2$ as
\begin{equation}
\Phi_0 = \frac{1}{T \pi \theta_1^2 \cdot (1-r^2)} 
(\frac{N_2}{r^2} - r^2 N_1 )
\end{equation}
We note that the neutrino observed by the virtual detector II is
also observed by virtual detector I, therefore,
$N_1 - N_2$ and $N_2$ are independent variables 
rather than $N_1$ and $N_2$.
Therefore, in terms of independent variables,
the ``true'' flux $\Phi_0$ is rewritten as,
\begin{equation}
\label{eq:point_flux2}
\Phi_0 = \frac{1}{T \pi \theta_1^2 \cdot (1-r^2)} 
\left((\frac{1}{r^2} -r^2)N_2 - r^2 (N_1-N_2)\right)
\end{equation}

\begin{figure}[htb]
  \centering{
  \includegraphics[height=2.2in]{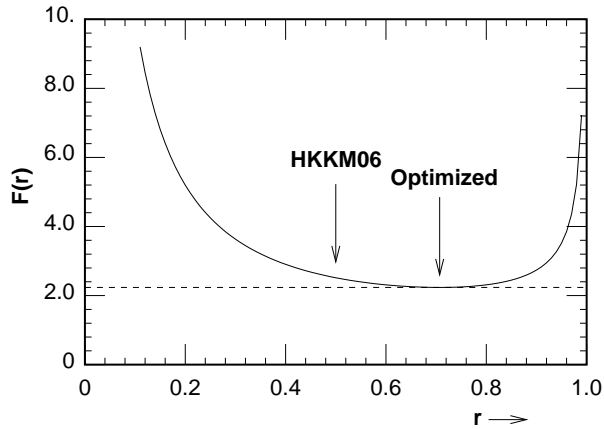}
  }\caption{
The variation of the function $F(r)$ defined by \ref{eq:error_point_flux8}.
It has a minimum value $\sqrt{5}\simeq 2.236$ at 
$r=1/\sqrt{2}\simeq 0.707$ indicated by ``Optimized''.
The ``HKKM06'' indicates the $r=0.5$ used in HKKM06, where the $F$-value is a little
larger than the optimized one ($F(0.5)=2.517$).
}
\label{virtual-ratio-statis}
\end{figure}

Assuming Poissonian fluctuations for the numbers of the observed  
neutrinos, we can estimate the statistical error of the ``true'' flux as
\begin{equation}
\label{eq:error_point_flux7}
\Delta\Phi_0
= \frac{1}{T \pi \theta_1^2 \cdot (1-r^2)}\sqrt{\left((\frac{1}{r^2}-r^2)^2r^2+ r^4(1-r^2)\right) N_1}~~,
\end{equation}
where we used the approximate relations $N_2 \simeq r^2 N_1$ and 
$N_1- N_2 \simeq (1-r^2)N_1$.
Therefore the ratio of $\Delta\Phi_0/\Delta\Phi_1$ is given as
\begin{equation}
\label{eq:error_point_flux8}
\frac{\Delta\Phi_0} {\Delta\Phi_1} 
= \frac{1}{1-r^2}\sqrt{(\frac{1}{r^2}-r^2)^2r^2+ r^4(1-r^2)}
\equiv F(r) ~~.
\end{equation}
The function $F(r)$ has a minimum at $r=1/\sqrt{2}\simeq 0.707$ and 
$F(1/\sqrt{2})=\sqrt(5)\simeq 2.236$.
We take the radius of the virtual detector I as 1113.6\,km corresponding to
an angle of 10$^\circ$  at the center of the Earth, and 
the radius of the virtual detector II as 787.4\,km corresponding to 
7.071$^\circ$.

The ``virtual detector correction'' is studied in the [before$/$after] ratio
in our calculation for Kamioka in Fig.~\ref{virtual-ratio-statis},
for $1>\cos\theta_z>0.9$ as the near vertical directions
and for
$0.1>\cos\theta_z>0$ as the near horizontal directions,
where $\theta_z$ is the arrival zenith angle.
Note,  the ``virtual detector correction'' is necessary 
and effective for the downward going neutrinos~\cite{hkkms2006}.

\begin{figure}[htb]
  \centering{
  \includegraphics[height=2.2in]{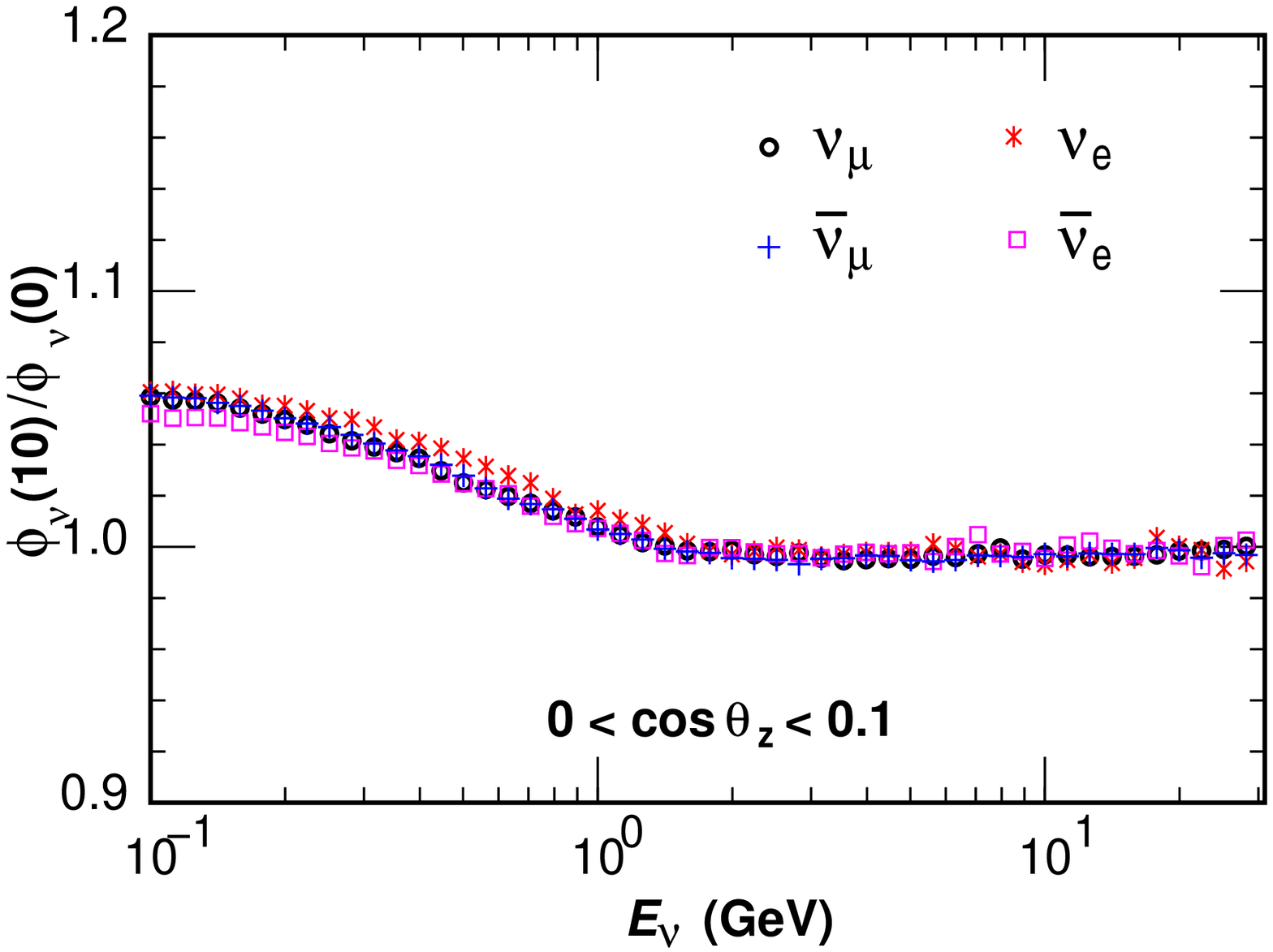}
  \includegraphics[height=2.2in]{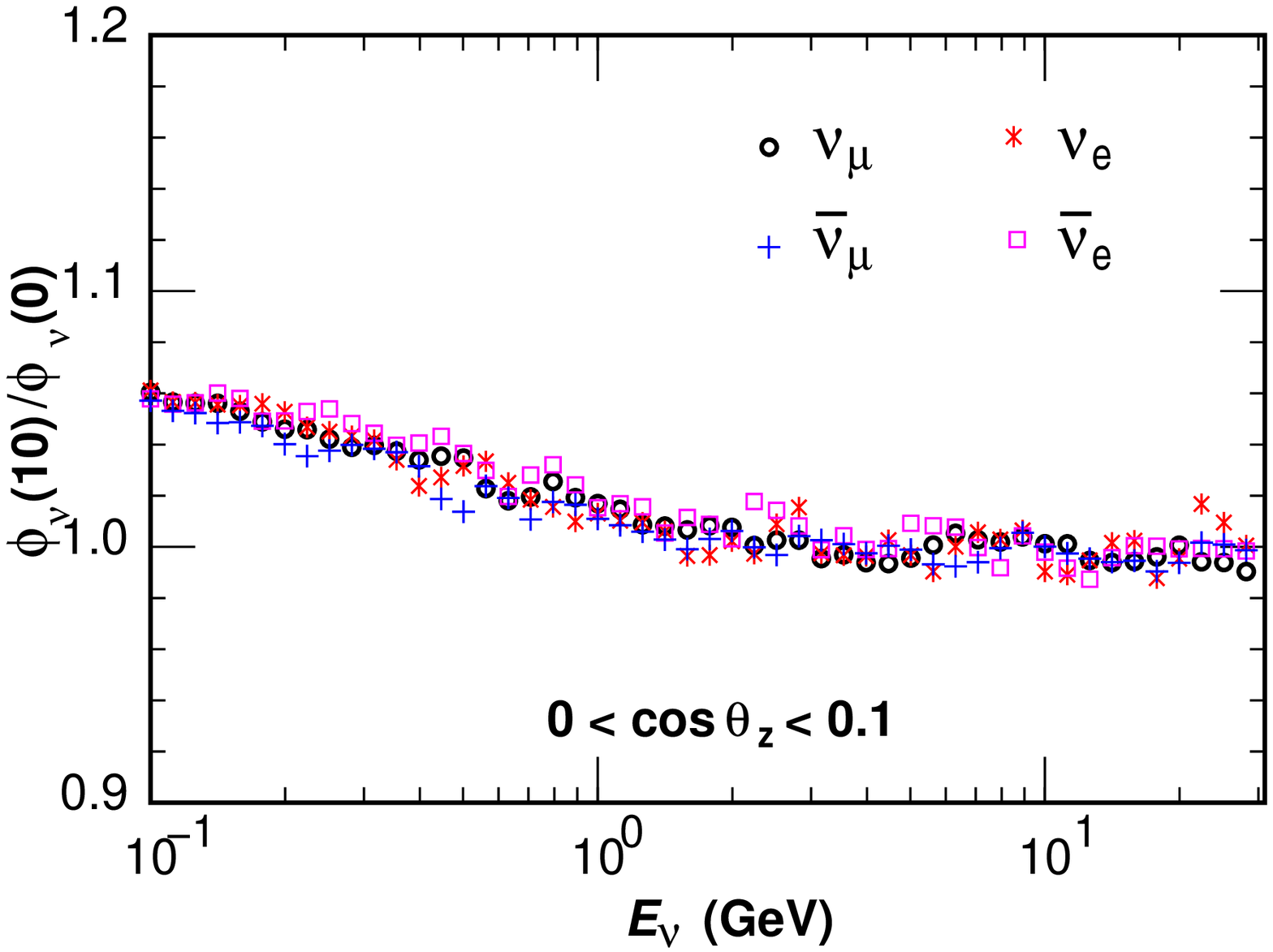}
  }\caption{
Ratio of the atmospheric neutrino flux averaged over the virtual 
detector ($\phi(10)$)  to 
the flux for the true neutrino detector ($\phi(0)$) estimated with the 
procedure explained in the text. The atmospheric neutrino fluxes are
averaged over all the azimuth angles.
Here, $\theta_z$ stands for the zenith angle of the arrival direction, and
the left panel is for the near vertical downward directions, and the right 
panel for the near horizontal directions.
}
\label{virtual-ratio-aa}
\end{figure}

\begin{figure}[htb]
\centering{
\includegraphics[height=2.2in]{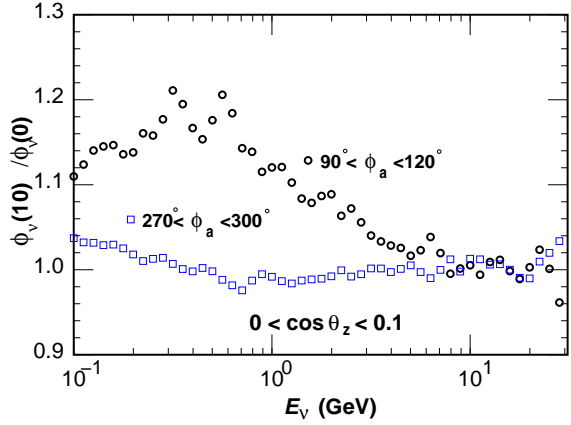}
 }\caption{
Ratio of the neutrino fluxes averaged over the virtual detector ($\phi(10)$)  
to 
the flux for the true neutrino detector ($\phi(0)$) calculated with the 
procedure explained in the text for near horizontal east 
($90^\circ < \phi_a < 120^\circ$) directions.
}
  \label{virtual-ratio-ew}
\end{figure}

From the figures, we find the ``virtual detector correction'' is 
needed below 3~GeV even for the average over all azimuth directions.
When we divide all the arrival azimuth angles into 12 bins, we
find the largest correction in the azimuth bin of
$[90^\circ, 120^\circ]$ for near horizontal directions ($0.1>\cos\theta_z>0$).
Note, we measure the azimuth angle counterclockwise from south.
We show the [before$/$after] ratio of the flux for this bin
 and opposite azimuth bin ([$270^\circ,300^\circ$]) 
in Fig.~\ref{virtual-ratio-ew}, summing over all kinds of neutrino.

\section{\label{statistical-error}
Analysis of simulation data and the statistical error
}

To analyze the simulation output, we take a logarithmically 
equally-spaced grid {$E_i$} for the neutrino energy as
\begin{equation}
E_i = E_0 \cdot R^{i}, ~~~~(i= . ., -1, 0, 1, ....),
\end{equation}
and count the the numbers of neutrinos falling in each 
[$E_{i+1}, E_{i}$]
bin for each kind of neutrino and each arrival direction window.
Note, we take $E_0$ = 0.1~GeV and $R=10^{1/20}$, then used grids
with $i \ge -1$ in this study.
As usual, we assume that the integration of the neutrino flux 
in a energy bin is related to the number of neutrino in the bin as
\begin{equation}
\label{eq:num2flux}
I_i = \int_{E_{i}}^{E_{i+1}}  \Phi(E) dE 
= \frac{N_i}{S\cdot \Omega \cdot T}
\end{equation}
where $N_i$ is the number of neutrinos in the [$E_{i+1}, E_{i}$] bin,
$T$ is the equivalent time of the simulation, $\Omega$ is the opening angle 
of the window, and $S$ is the area of the virtual detector.

We calculate the neutrino flux $\Phi_i$ at energy $E_i$ from 
$I_{i}$ and $I_{i-1}$
which are integral fluxes calculated from the simulation
in the two adjacent energy bins [$E_{i+1}, E_{i}$] and [$E_{i}, E_{i-1}$]
respectively.
For these two energy bins, we approximate the neutrino flux as the power 
function of the energy as $\Phi(E) = A \cdot E^\alpha$, 
and  $\Phi_i$ is calculated as
\begin{equation}
\label{eq:2bin-flux}
\Phi_i = \frac{I_i I_{i-1} \left( \ln(I_i) -\ln(I_{i-1}) \right)}
{ E_i \ln(R) \left(I_i - I_{i-1}\right)}~~~~.
\end{equation}
This expression is singular at $I_i = I_{i-1}$,
but it becomes a smooth function of $I_i$ and $I_{i-1}$, if we 
define the value at $I_i = I_{i-1}$ as
\begin{equation}
\label{eq:2bin-flux2}
\Phi_i = \frac{I_i}{ E_i \ln(R)}~~~~,
\end{equation}
to which 
~\ref{eq:2bin-flux} approaches in the $I_i \rightarrow I_{i-1}$ limit.

Assuming Poissonian fluctuations for the statistical error,
the error for integrated flux $\Delta I_i/I_i$ is given by $1/\sqrt{N_i}$.
The statistical error for $\Phi_i$ is estimated substituting 
~\ref{eq:2bin-flux} into the general expression
\begin{equation}
\label{eq:2bin-error}
\Delta \Phi_i = \sqrt{
 \left ({\frac{\partial \Phi_i} {\partial I_i}}\Delta I_i \right )^2 + 
\left ({\frac{\partial \Phi_i} {\partial I_{i-1}}}\Delta I_{i-1} \right )^2
}~~,
\end{equation}
and the ratio $\Delta \Phi_i/\Phi_i$ is given as

\begin{equation}
\label{eq:2bin-error3}
\frac{\Delta \Phi_i}{\Phi_i} = \sqrt{
\left ({\frac{1} {\ln x}} - {\frac{1}{x-1}} \right)^2
\frac{1}{N_i} + 
\left (
 \frac{1}{\ln (1/x)} - \frac{1}{1/x -1} \right)^2
\frac{1}{N_{i-1}} 
}~~~,
\end{equation}
where $x=I_i/I_{i-1}=N_i/N_{i-1}$. 
This expression is well approximated by the simpler relation,
\begin{equation}
\label{eq:2bin-error4}
\frac{\Delta \Phi_i}{\Phi_i} \simeq \frac{1}{2}\sqrt{
\frac{1}{N_i} + 
\frac{1}{N_{i-1}} 
}~~~.
\end{equation}
In the $N_i \rightarrow N_{i-1}$ limit,
both \ref{eq:2bin-error3} and \ref{eq:2bin-error4} give $1/\sqrt{N_1+N_2}$, 
and the differences between them are less than 5~\% for $0.5 < N_i / N_{i-1} < 2$.

\section{\label{nflx-tbl}Atmospheric neutrino flux below 32 GeV}

Here we tabulate the 
atmospheric neutrino flux for solar minimum at sea level
at locations --- Kamioka, Sudbury (North America), 
and Gran Sasso --- as in the former publication,
averaging them over all the arrival azimuth directions and 
in the zenith angle bins with $\Delta \cos\theta_z = 0.1$,
where $\theta_z$ is the zenith angle of the arrival direction
of the neutrino, in Tables I--XX.

Due to page limitations, we cannot present all the
calculated results here. The atmospheric neutrino fluxes for
different solar activity  for Kamioka, Sudbury, Soudan2 site, 
Gran Sasso, Frejus, and Homestake will be available at the web site:
\url{http://www.icrr.u-tokyo.ac.jp/~mhonda}.

\begin{table}[tbh]
\caption{Neutrino flux ($\rm m^{-2}sec^{-1}sr^{-1}GeV^{-1}$) for 
$1.0 \ge \cos\theta_z > 0.9$}
\begin{tabular}{c | r r r r| r r r r| r r r r | c}
\toprule
\multicolumn{1}{c}{}&\multicolumn{4}{|c}{Kamioka}&\multicolumn{4}{|c}{Sudbury}
&\multicolumn{4}{|c}{Gran Sasso}&\multicolumn{1}{|c}{}\\
$E_\nu$ (GeV)&$\nu_\mu$~~&$\bar\nu_\mu$~~&$\nu_e$~~&$\bar\nu_e$~~&$\nu_\mu$~~&$\bar\nu_\mu$~~&$\nu_e$~~&$\bar\nu_e$~~&$\nu_\mu$~~&$\bar\nu_\mu$~~&$\nu_e$~~&$\bar\nu_e$~~&~$Norm$~ ~\\
\toprule
0.1000&  4.25&  4.25&  2.02&  2.08& 11.91& 12.05&  6.01&  5.34&  7.34&  7.43&  3.56&  3.51& $10^{ 3}$\\
0.1259&  3.52&  3.52&  1.66&  1.68&  9.53&  9.58&  4.76&  4.17&  6.01&  6.08&  2.90&  2.81& $10^{ 3}$\\
0.1585&  2.81&  2.81&  1.31&  1.31&  7.27&  7.27&  3.60&  3.09&  4.74&  4.77&  2.26&  2.14& $10^{ 3}$\\
0.1995& 21.50& 21.50&  9.98&  9.81& 52.12& 51.97& 25.95& 21.84& 35.54& 35.52& 16.86& 15.69& $10^{ 2}$\\
0.2512& 15.68& 15.68&  7.32&  7.04& 35.16& 34.99& 17.80& 14.63& 25.33& 25.16& 12.03& 10.97& $10^{ 2}$\\
0.3162& 11.02& 11.02&  5.13&  4.84& 22.53& 22.42& 11.51&  9.32& 17.23& 17.02&  8.21&  7.30& $10^{ 2}$\\
\botrule
\end{tabular}
\end{table}
 
\begin{table}[tbh]
\caption{Neutrino flux ($\rm m^{-2}sec^{-1}sr^{-1}GeV^{-1}$) for $0.9 \ge \cos\theta_z > 0.8$}
\begin{tabular}{c | r r r r| r r r r| r r r r | c }
\toprule
\multicolumn{1}{c}{}&\multicolumn{4}{|c}{Kamioka}&\multicolumn{4}{|c}{Sudbury}&\multicolumn{4}{|c}{Gran Sasso}&\multicolumn{1}{|c}{}\\
$E_\nu$ (GeV)&$\nu_\mu$~~&$\bar\nu_\mu$~~&$\nu_e$~~&$\bar\nu_e$~~&$\nu_\mu$~~&$\bar\nu_\mu$~~&$\nu_e$~~&$\bar\nu_e$~~&$\nu_\mu$~~&$\bar\nu_\mu$~~&$\nu_e$~~&$\bar\nu_e$~~&$Norm$~ \\
\toprule
0.1000&  4.33&  4.33&  2.06&  2.11& 12.25& 12.38&  6.17&  5.50&  7.50&  7.61&  3.65&  3.59& $10^{ 3}$\\
0.1259&  3.57&  3.57&  1.69&  1.71&  9.71&  9.76&  4.87&  4.27&  6.12&  6.18&  2.95&  2.86& $10^{ 3}$\\
0.1585&  2.84&  2.84&  1.33&  1.33&  7.37&  7.38&  3.68&  3.17&  4.81&  4.83&  2.30&  2.19& $10^{ 3}$\\
0.1995& 21.65& 21.65& 10.20&  9.94& 52.72& 52.62& 26.52& 22.35& 35.98& 35.94& 17.24& 15.99& $10^{ 2}$\\
0.2512& 15.80& 15.80&  7.47&  7.17& 35.57& 35.44& 18.22& 15.02& 25.60& 25.49& 12.32& 11.24& $10^{ 2}$\\
0.3162& 11.09& 11.09&  5.27&  4.95& 22.84& 22.72& 11.84&  9.59& 17.48& 17.27&  8.45&  7.51& $10^{ 2}$\\
\botrule
\end{tabular}
\end{table} 
 
\begin{table}[tbh]
\caption{Neutrino flux ($\rm m^{-2}sec^{-1}sr^{-1}GeV^{-1}$) for $0.8 \ge \cos\theta_z> 0.7$}
\begin{tabular}{c | r r r r| r r r r| r r r r | c }
\toprule
\multicolumn{1}{c}{}&\multicolumn{4}{|c}{Kamioka}&\multicolumn{4}{|c}{Sudbury }&\multicolumn{4}{|c}{Gran Sasso}&\multicolumn{1}{|c}{}\\
$E_\nu$ (GeV)&$\nu_\mu$~~&$\bar\nu_\mu$~~&$\nu_e$~~&$\bar\nu_e$~~&$\nu_\mu$~~&$\bar\nu_\mu$~~&$\nu_e$~~&$\bar\nu_e$~~&$\nu_\mu$~~&$\bar\nu_\mu$~~&$\nu_e$~~&$\bar\nu_e$~~&$Norm$~ \\
\toprule
0.1000&  4.46&  4.46&  2.12&  2.17& 12.68& 12.82&  6.38&  5.70&  7.74&  7.84&  3.75&  3.70& $10^{ 3}$\\
0.1259&  3.63&  3.63&  1.73&  1.74&  9.97& 10.02&  5.01&  4.40&  6.25&  6.31&  3.03&  2.94& $10^{ 3}$\\
0.1585&  2.88&  2.88&  1.36&  1.35&  7.50&  7.51&  3.77&  3.24&  4.88&  4.90&  2.35&  2.24& $10^{ 3}$\\
0.1995&  2.18&  2.18&  1.04&  1.01&  5.35&  5.33&  2.71&  2.29&  3.64&  3.64&  1.76&  1.64& $10^{ 3}$\\
0.2512& 15.92& 15.92&  7.63&  7.30& 36.02& 35.91& 18.63& 15.42& 25.92& 25.74& 12.63& 11.50& $10^{ 2}$\\
0.3162& 11.19& 11.19&  5.41&  5.07& 23.15& 23.10& 12.18&  9.89& 17.65& 17.48&  8.69&  7.72& $10^{ 2}$\\
\botrule
\end{tabular}
\end{table} 
 
\begin{table}[tbh]
\caption{Neutrino flux ($\rm m^{-2}sec^{-1}sr^{-1}GeV^{-1}$) for $0.7 \ge \cos\theta_z> 0.6$}
\begin{tabular}{c | r r r r| r r r r| r r r r | c }
\toprule
\multicolumn{1}{c}{}&\multicolumn{4}{|c}{Kamioka}&\multicolumn{4}{|c}{Sudbury }&\multicolumn{4}{|c}{Gran Sasso}&\multicolumn{1}{|c}{}\\
$E_\nu$ (GeV)&$\nu_\mu$~~&$\bar\nu_\mu$~~&$\nu_e$~~&$\bar\nu_e$~~&$\nu_\mu$~~&$\bar\nu_\mu$~~&$\nu_e$~~&$\bar\nu_e$~~&$\nu_\mu$~~&$\bar\nu_\mu$~~&$\nu_e$~~&$\bar\nu_e$~~&$Norm$~ \\
\toprule
0.1000&  4.63&  4.63&  2.20&  2.24& 13.31& 13.43&  6.67&  5.99&  8.05&  8.17&  3.90&  3.85& $10^{ 3}$\\
0.1259&  3.74&  3.74&  1.78&  1.79& 10.31& 10.36&  5.20&  4.58&  6.44&  6.50&  3.12&  3.03& $10^{ 3}$\\
0.1585&  2.94&  2.94&  1.40&  1.38&  7.70&  7.70&  3.88&  3.35&  4.98&  5.02&  2.41&  2.29& $10^{ 3}$\\
0.1995&  2.22&  2.22&  1.06&  1.03&  5.45&  5.44&  2.79&  2.36&  3.70&  3.69&  1.80&  1.68& $10^{ 3}$\\
0.2512& 16.09& 16.09&  7.82&  7.43& 36.61& 36.53& 19.16& 15.84& 26.20& 26.11& 12.95& 11.77& $10^{ 2}$\\
0.3162& 11.28& 11.28&  5.54&  5.17& 23.48& 23.47& 12.48& 10.20& 17.87& 17.71&  8.92&  7.93& $10^{ 2}$\\
\botrule
\end{tabular}
\end{table} 
 
\begin{table}[tbh]
\caption{Neutrino flux ($\rm m^{-2}sec^{-1}sr^{-1}GeV^{-1}$) for $0.6 \ge \cos\theta_z> 0.5$}
\begin{tabular}{c | r r r r| r r r r| r r r r | c }
\toprule
\multicolumn{1}{c}{}&\multicolumn{4}{|c}{Kamioka}&\multicolumn{4}{|c}{Sudbury }&\multicolumn{4}{|c}{Gran Sasso}&\multicolumn{1}{|c}{}\\
$E_\nu$ (GeV)&$\nu_\mu$~~&$\bar\nu_\mu$~~&$\nu_e$~~&$\bar\nu_e$~~&$\nu_\mu$~~&$\bar\nu_\mu$~~&$\nu_e$~~&$\bar\nu_e$~~&$\nu_\mu$~~&$\bar\nu_\mu$~~&$\nu_e$~~&$\bar\nu_e$~~&$Norm$~ \\
\toprule
0.1000&  4.88&  4.88&  2.31&  2.35& 14.19& 14.33&  7.11&  6.38&  8.53&  8.65&  4.11&  4.06& $10^{ 3}$\\
0.1259&  3.89&  3.89&  1.86&  1.85& 10.84& 10.90&  5.47&  4.84&  6.72&  6.79&  3.26&  3.17& $10^{ 3}$\\
0.1585&  3.03&  3.03&  1.45&  1.42&  8.00&  8.00&  4.05&  3.50&  5.15&  5.17&  2.50&  2.37& $10^{ 3}$\\
0.1995&  2.27&  2.27&  1.09&  1.05&  5.61&  5.60&  2.89&  2.45&  3.79&  3.78&  1.85&  1.73& $10^{ 3}$\\
0.2512& 16.35& 16.35&  8.02&  7.58& 37.43& 37.37& 19.71& 16.38& 26.66& 26.53& 13.28& 12.09& $10^{ 2}$\\
0.3162& 11.43& 11.43&  5.68&  5.29& 23.93& 23.87& 12.90& 10.52& 18.10& 17.94&  9.14&  8.15& $10^{ 2}$\\
\botrule
\end{tabular}
\end{table} 
 
\begin{table}[tbh]
\caption{Neutrino flux ($\rm m^{-2}sec^{-1}sr^{-1}GeV^{-1}$) for $0.5 \ge \cos\theta_z > 0.4$}
\begin{tabular}{c | r r r r| r r r r| r r r r | c }
\toprule
\multicolumn{1}{c}{}&\multicolumn{4}{|c}{Kamioka}&\multicolumn{4}{|c}{Sudbury }&\multicolumn{4}{|c}{Gran Sasso}&\multicolumn{1}{|c}{}\\
$E_\nu$ (GeV)&$\nu_\mu$~~&$\bar\nu_\mu$~~&$\nu_e$~~&$\bar\nu_e$~~&$\nu_\mu$~~&$\bar\nu_\mu$~~&$\nu_e$~~&$\bar\nu_e$~~&$\nu_\mu$~~&$\bar\nu_\mu$~~&$\nu_e$~~&$\bar\nu_e$~~&$Norm$~ \\
\toprule
0.1000&  5.28&  5.28&  2.49&  2.52& 15.55& 15.71&  7.75&  6.99&  9.26&  9.39&  4.45&  4.40& $10^{ 3}$\\
0.1259&  4.15&  4.15&  1.97&  1.97& 11.65& 11.72&  5.89&  5.22&  7.18&  7.25&  3.48&  3.39& $10^{ 3}$\\
0.1585&  3.17&  3.17&  1.53&  1.48&  8.45&  8.46&  4.30&  3.73&  5.42&  5.44&  2.63&  2.50& $10^{ 3}$\\
0.1995&  2.35&  2.35&  1.14&  1.09&  5.86&  5.84&  3.03&  2.57&  3.93&  3.92&  1.93&  1.80& $10^{ 3}$\\
0.2512& 16.79& 16.79&  8.31&  7.81& 38.68& 38.57& 20.57& 17.12& 27.36& 27.26& 13.73& 12.54& $10^{ 2}$\\
0.3162& 11.64& 11.64&  5.86&  5.41& 24.52& 24.50& 13.39& 10.96& 18.40& 18.27&  9.41&  8.39& $10^{ 2}$\\
\botrule
\end{tabular}
\end{table} 
 
\begin{table}[tbh]
\caption{Neutrino flux ($\rm m^{-2}sec^{-1}sr^{-1}GeV^{-1}$) for $0.4 \ge \cos\theta_z > 0.3$}
\begin{tabular}{c | r r r r| r r r r| r r r r | c }
\toprule
\multicolumn{1}{c}{}&\multicolumn{4}{|c}{Kamioka}&\multicolumn{4}{|c}{Sudbury }&\multicolumn{4}{|c}{Gran Sasso}&\multicolumn{1}{|c}{}\\
$E_\nu$ (GeV)&$\nu_\mu$~~&$\bar\nu_\mu$~~&$\nu_e$~~&$\bar\nu_e$~~&$\nu_\mu$~~&$\bar\nu_\mu$~~&$\nu_e$~~&$\bar\nu_e$~~&$\nu_\mu$~~&$\bar\nu_\mu$~~&$\nu_e$~~&$\bar\nu_e$~~&$Norm$~ \\
\toprule
0.1000&  5.93&  5.93&  2.78&  2.80& 17.73& 17.92&  8.82&  7.96& 10.45& 10.62&  4.99&  4.95& $10^{ 3}$\\
0.1259&  4.56&  4.56&  2.18&  2.15& 13.02& 13.08&  6.59&  5.85&  7.94&  8.02&  3.84&  3.74& $10^{ 3}$\\
0.1585&  3.44&  3.44&  1.65&  1.60&  9.26&  9.27&  4.72&  4.11&  5.88&  5.91&  2.86&  2.73& $10^{ 3}$\\
0.1995&  2.49&  2.49&  1.22&  1.16&  6.30&  6.29&  3.27&  2.79&  4.19&  4.19&  2.07&  1.93& $10^{ 3}$\\
0.2512& 17.56& 17.56&  8.77&  8.18& 40.80& 40.89& 21.94& 18.33& 28.67& 28.53& 14.48& 13.24& $10^{ 2}$\\
0.3162& 12.01& 12.01&  6.17&  5.60& 25.55& 25.57& 14.10& 11.60& 19.05& 18.83&  9.84&  8.77& $10^{ 2}$\\
\botrule
\end{tabular}
\end{table} 
 
\begin{table}[tbh]
\caption{Neutrino flux ($\rm m^{-2}sec^{-1}sr^{-1}GeV^{-1}$) for $0.3 \ge \cos\theta_z > 0.2$}
\begin{tabular}{c | r r r r| r r r r| r r r r | c }
\toprule
\multicolumn{1}{c}{}&\multicolumn{4}{|c}{Kamioka}&\multicolumn{4}{|c}{Sudbury }&\multicolumn{4}{|c}{Gran Sasso}&\multicolumn{1}{|c}{}\\
$E_\nu$ (GeV)&$\nu_\mu$~~&$\bar\nu_\mu$~~&$\nu_e$~~&$\bar\nu_e$~~&$\nu_\mu$~~&$\bar\nu_\mu$~~&$\nu_e$~~&$\bar\nu_e$~~&$\nu_\mu$~~&$\bar\nu_\mu$~~&$\nu_e$~~&$\bar\nu_e$~~&$Norm$~ \\
\toprule
0.1000&  7.15&  7.15&  3.32&  3.36& 21.70& 21.92& 10.74&  9.74& 12.65& 12.85&  6.00&  5.98& $10^{ 3}$\\
0.1259&  5.36&  5.36&  2.55&  2.52& 15.49& 15.62&  7.86&  7.01&  9.37&  9.46&  4.52&  4.44& $10^{ 3}$\\
0.1585&  3.95&  3.95&  1.90&  1.83& 10.79& 10.79&  5.51&  4.81&  6.79&  6.81&  3.30&  3.15& $10^{ 3}$\\
0.1995&  2.80&  2.80&  1.38&  1.30&  7.16&  7.13&  3.75&  3.21&  4.72&  4.70&  2.34&  2.18& $10^{ 3}$\\
0.2512& 19.25& 19.25&  9.74&  8.95& 45.31& 45.26& 24.59& 20.55& 31.50& 31.30& 16.08& 14.67& $10^{ 2}$\\
0.3162& 12.89& 12.89&  6.68&  6.04& 27.75& 27.67& 15.46& 12.78& 20.40& 20.22& 10.68&  9.50& $10^{ 2}$\\
\botrule
\end{tabular}
\end{table} 
 
\begin{table}[tbh]
\caption{Neutrino flux ($\rm m^{-2}sec^{-1}sr^{-1}GeV^{-1}$) for $0.2 \ge \cos\theta_z > 0.1$}
\begin{tabular}{c | r r r r| r r r r| r r r r | c }
\toprule
\multicolumn{1}{c}{}&\multicolumn{4}{|c}{Kamioka}&\multicolumn{4}{|c}{Sudbury }&\multicolumn{4}{|c}{Gran Sasso}&\multicolumn{1}{|c}{}\\
$E_\nu$ (GeV)&$\nu_\mu$~~&$\bar\nu_\mu$~~&$\nu_e$~~&$\bar\nu_e$~~&$\nu_\mu$~~&$\bar\nu_\mu$~~&$\nu_e$~~&$\bar\nu_e$~~&$\nu_\mu$~~&$\bar\nu_\mu$~~&$\nu_e$~~&$\bar\nu_e$~~&$Norm$~ \\
\toprule
0.1000&  9.80&  9.80&  4.53&  4.53& 29.97& 30.35& 14.81& 13.52& 17.36& 17.68&  8.17&  8.18& $10^{ 3}$\\
0.1259&  7.14&  7.14&  3.39&  3.32& 20.86& 20.98& 10.57&  9.47& 12.52& 12.61&  6.02&  5.92& $10^{ 3}$\\
0.1585&  5.08&  5.08&  2.45&  2.35& 14.07& 14.11&  7.23&  6.34&  8.78&  8.82&  4.27&  4.09& $10^{ 3}$\\
0.1995&  3.51&  3.51&  1.73&  1.62&  9.06&  9.04&  4.78&  4.09&  5.92&  5.92&  2.95&  2.75& $10^{ 3}$\\
0.2512&  2.34&  2.34&  1.19&  1.09&  5.57&  5.56&  3.05&  2.57&  3.83&  3.79&  1.97&  1.79& $10^{ 3}$\\
0.3162& 15.13& 15.13&  7.94&  7.04& 32.95& 32.93& 18.66& 15.46& 23.90& 23.76& 12.75& 11.32& $10^{ 2}$\\
\botrule
\end{tabular}
\end{table}

\begin{table}[tbh]
\caption{Neutrino flux ($\rm m^{-2}sec^{-1}sr^{-1}GeV^{-1}$) for $0.1 \ge \cos\theta_z> 0.0$}
\begin{tabular}{c | r r r r| r r r r| r r r r | c }
\toprule
\multicolumn{1}{c}{}&\multicolumn{4}{|c}{Kamioka}&\multicolumn{4}{|c}{Sudbury }&\multicolumn{4}{|c}{Gran Sasso}&\multicolumn{1}{|c}{}\\
$E_\nu$ (GeV)&$\nu_\mu$~~&$\bar\nu_\mu$~~&$\nu_e$~~&$\bar\nu_e$~~&$\nu_\mu$~~&$\bar\nu_\mu$~~&$\nu_e$~~&$\bar\nu_e$~~&$\nu_\mu$~~&$\bar\nu_\mu$~~&$\nu_e$~~&$\bar\nu_e$~~&$Norm$~ \\
\toprule
0.1000& 15.94& 15.94&  7.20&  7.32& 48.03& 48.70& 23.42& 21.70& 28.03& 28.57& 13.04& 13.14& $10^{ 3}$\\
0.1259& 11.21& 11.21&  5.25&  5.20& 32.39& 32.64& 16.36& 14.78& 19.52& 19.71&  9.35&  9.21& $10^{ 3}$\\
0.1585&  7.70&  7.70&  3.68&  3.53& 21.19& 21.31& 10.82&  9.64& 13.23& 13.31&  6.45&  6.20& $10^{ 3}$\\
0.1995&  5.13&  5.13&  2.51&  2.36& 13.19& 13.19&  6.97&  6.02&  8.61&  8.56&  4.30&  4.01& $10^{ 3}$\\
0.2512&  3.27&  3.27&  1.67&  1.51&  7.79&  7.81&  4.31&  3.63&  5.34&  5.31&  2.78&  2.51& $10^{ 3}$\\
0.3162& 20.11& 20.11& 10.78&  9.48& 44.46& 44.36& 25.43& 21.20& 31.96& 31.86& 17.31& 15.26& $10^{ 2}$\\
\botrule
\end{tabular}
\end{table} 
 
\begin{table}[tbh]
\caption{Neutrino flux ($\rm m^{-2}sec^{-1}sr^{-1}GeV^{-1}$) for $0.0 \ge \cos\theta_z> -0.1$}
\begin{tabular}{c | r r r r| r r r r| r r r r | c }
\toprule
\multicolumn{1}{c}{}&\multicolumn{4}{|c}{Kamioka}&\multicolumn{4}{|c}{Sudbury }&\multicolumn{4}{|c}{Gran Sasso}&\multicolumn{1}{|c}{}\\
$E_\nu$ (GeV)&$\nu_\mu$~~&$\bar\nu_\mu$~~&$\nu_e$~~&$\bar\nu_e$~~&$\nu_\mu$~~&$\bar\nu_\mu$~~&$\nu_e$~~&$\bar\nu_e$~~&$\nu_\mu$~~&$\bar\nu_\mu$~~&$\nu_e$~~&$\bar\nu_e$~~&$Norm$~ \\
\toprule
0.1000& 17.11& 17.11&  7.76&  7.86& 46.20& 46.96& 22.44& 20.95& 28.33& 28.90& 13.36& 13.18& $10^{ 3}$\\
0.1259& 11.97& 11.97&  5.60&  5.59& 31.24& 31.62& 15.76& 14.38& 19.72& 19.87&  9.46&  9.26& $10^{ 3}$\\
0.1585&  8.18&  8.18&  3.92&  3.78& 20.60& 20.68& 10.48&  9.40& 13.29& 13.37&  6.53&  6.18& $10^{ 3}$\\
0.1995&  5.36&  5.36&  2.65&  2.49& 12.92& 12.89&  6.80&  5.92&  8.58&  8.55&  4.33&  3.98& $10^{ 3}$\\
0.2512&  3.38&  3.38&  1.73&  1.59&  7.68&  7.67&  4.21&  3.59&  5.30&  5.28&  2.78&  2.50& $10^{ 3}$\\
0.3162& 20.65& 20.65& 11.17&  9.85& 44.06& 44.13& 25.15& 21.03& 31.55& 31.53& 17.27& 15.00& $10^{ 2}$\\
\botrule
\end{tabular}
\end{table} 

\begin{table}[tbh]
\caption{Neutrino flux ($\rm m^{-2}sec^{-1}sr^{-1}GeV^{-1}$) for $-0.1 \ge \cos\theta_z > -0.2$}
\begin{tabular}{c | r r r r| r r r r| r r r r | c }
\toprule
\multicolumn{1}{c}{}&\multicolumn{4}{|c}{Kamioka}&\multicolumn{4}{|c}{Sudbury }&\multicolumn{4}{|c}{Gran Sasso}&\multicolumn{1}{|c}{}\\
$E_\nu$ (GeV)&$\nu_\mu$~~&$\bar\nu_\mu$~~&$\nu_e$~~&$\bar\nu_e$~~&$\nu_\mu$~~&$\bar\nu_\mu$~~&$\nu_e$~~&$\bar\nu_e$~~&$\nu_\mu$~~&$\bar\nu_\mu$~~&$\nu_e$~~&$\bar\nu_e$~~&$Norm$~ \\
\toprule
0.1000& 12.56& 12.56&  5.81&  5.81& 25.56& 25.94& 12.51& 11.63& 17.71& 18.04&  8.58&  8.18& $10^{ 3}$\\
0.1259&  8.94&  8.94&  4.27&  4.18& 17.98& 18.12&  9.03&  8.25& 12.60& 12.72&  6.26&  5.84& $10^{ 3}$\\
0.1585&  6.24&  6.24&  3.04&  2.89& 12.30& 12.38&  6.24&  5.59&  8.69&  8.78&  4.36&  3.99& $10^{ 3}$\\
0.1995&  4.18&  4.18&  2.09&  1.94&  8.07&  8.06&  4.20&  3.67&  5.77&  5.80&  2.97&  2.65& $10^{ 3}$\\
0.2512&  2.69&  2.69&  1.40&  1.27&  5.05&  5.05&  2.72&  2.33&  3.67&  3.68&  1.95&  1.71& $10^{ 3}$\\
0.3162& 16.89& 16.89&  9.05&  8.03& 30.64& 30.52& 17.09& 14.31& 22.59& 22.71& 12.46& 10.64& $10^{ 2}$\\
\botrule
\end{tabular}
\end{table} 

\begin{table}[tbh]
\caption{Neutrino flux ($\rm m^{-2}sec^{-1}sr^{-1}GeV^{-1}$) for $-0.2 \ge \cos\theta_z > -0.3$}
\begin{tabular}{c | r r r r| r r r r| r r r r | c }
\toprule
\multicolumn{1}{c}{}&\multicolumn{4}{|c}{Kamioka}&\multicolumn{4}{|c}{Sudbury }&\multicolumn{4}{|c}{Gran Sasso}&\multicolumn{1}{|c}{}\\
$E_\nu$ (GeV)&$\nu_\mu$~~&$\bar\nu_\mu$~~&$\nu_e$~~&$\bar\nu_e$~~&$\nu_\mu$~~&$\bar\nu_\mu$~~&$\nu_e$~~&$\bar\nu_e$~~&$\nu_\mu$~~&$\bar\nu_\mu$~~&$\nu_e$~~&$\bar\nu_e$~~&$Norm$~ \\
\toprule
0.1000& 10.41& 10.41&  4.95&  4.75& 15.38& 15.62&  7.58&  7.05& 12.45& 12.68&  6.14&  5.71& $10^{ 3}$\\
0.1259&  7.55&  7.55&  3.68&  3.48& 11.19& 11.33&  5.63&  5.15&  9.07&  9.19&  4.55&  4.16& $10^{ 3}$\\
0.1585&  5.34&  5.34&  2.64&  2.44&  7.93&  8.00&  4.01&  3.60&  6.44&  6.49&  3.26&  2.93& $10^{ 3}$\\
0.1995&  3.65&  3.65&  1.85&  1.67&  5.40&  5.42&  2.79&  2.45&  4.39&  4.41&  2.26&  2.00& $10^{ 3}$\\
0.2512&  2.39&  2.39&  1.25&  1.11&  3.53&  3.54&  1.88&  1.62&  2.88&  2.89&  1.53&  1.33& $10^{ 3}$\\
0.3162& 15.27& 15.27&  8.20&  7.15& 22.38& 22.41& 12.23& 10.35& 18.31& 18.37&  9.97&  8.50& $10^{ 2}$\\
\botrule
\end{tabular}
\end{table} 
 
\begin{table}[tbh]
\caption{Neutrino flux ($\rm m^{-2}sec^{-1}sr^{-1}GeV^{-1}$) for $-0.3 \ge \cos\theta_z > -0.4$}
\begin{tabular}{c | r r r r| r r r r| r r r r | c }
\toprule
\multicolumn{1}{c}{}&\multicolumn{4}{|c}{Kamioka}&\multicolumn{4}{|c}{Sudbury }&\multicolumn{4}{|c}{Gran Sasso}&\multicolumn{1}{|c}{}\\
$E_\nu$ (GeV)&$\nu_\mu$~~&$\bar\nu_\mu$~~&$\nu_e$~~&$\bar\nu_e$~~&$\nu_\mu$~~&$\bar\nu_\mu$~~&$\nu_e$~~&$\bar\nu_e$~~&$\nu_\mu$~~&$\bar\nu_\mu$~~&$\nu_e$~~&$\bar\nu_e$~~&$Norm$~ \\ 
\toprule
0.1000&  8.80&  8.80&  4.26&  4.01& 10.71& 10.89&  5.29&  4.91&  9.89& 10.05&  4.89&  4.54& $10^{ 3}$\\
0.1259&  6.53&  6.53&  3.21&  2.99&  7.99&  8.08&  4.00&  3.67&  7.37&  7.47&  3.70&  3.38& $10^{ 3}$\\
0.1585&  4.72&  4.72&  2.35&  2.15&  5.82&  5.87&  2.93&  2.65&  5.37&  5.41&  2.71&  2.42& $10^{ 3}$\\
0.1995&  3.29&  3.29&  1.67&  1.50&  4.08&  4.09&  2.09&  1.84&  3.74&  3.76&  1.92&  1.69& $10^{ 3}$\\
0.2512&  2.21&  2.21&  1.15&  1.01&  2.74&  2.74&  1.44&  1.25&  2.51&  2.52&  1.32&  1.14& $10^{ 3}$\\
0.3162& 14.38& 14.38&  7.64&  6.66& 17.81& 17.86&  9.55&  8.18& 16.28& 16.34&  8.74&  7.47& $10^{ 2}$\\
\botrule
\end{tabular}
\end{table} 
 
\begin{table}[tbh]
\caption{Neutrino flux ($\rm m^{-2}sec^{-1}sr^{-1}GeV^{-1}$) for $-0.4 \ge \cos\theta_z > -0.5$}
\begin{tabular}{c | r r r r| r r r r| r r r r | c }
\toprule
\multicolumn{1}{c}{}&\multicolumn{4}{|c}{Kamioka}&\multicolumn{4}{|c}{Sudbury }&\multicolumn{4}{|c}{Gran Sasso}&\multicolumn{1}{|c}{}\\
$E_\nu$ (GeV)&$\nu_\mu$~~&$\bar\nu_\mu$~~&$\nu_e$~~&$\bar\nu_e$~~&$\nu_\mu$~~&$\bar\nu_\mu$~~&$\nu_e$~~&$\bar\nu_e$~~&$\nu_\mu$~~&$\bar\nu_\mu$~~&$\nu_e$~~&$\bar\nu_e$~~&$Norm$~ \\
\toprule
0.1000&  7.89&  7.89&  3.83&  3.61&  8.34&  8.46&  4.11&  3.86&  8.52&  8.64&  4.22&  3.92& $10^{ 3}$\\
0.1259&  6.00&  6.00&  2.95&  2.75&  6.38&  6.45&  3.18&  2.94&  6.50&  6.57&  3.25&  2.98& $10^{ 3}$\\
0.1585&  4.43&  4.43&  2.20&  2.01&  4.74&  4.77&  2.37&  2.16&  4.82&  4.86&  2.42&  2.18& $10^{ 3}$\\
0.1995&  3.14&  3.14&  1.59&  1.43&  3.39&  3.39&  1.71&  1.53&  3.44&  3.45&  1.75&  1.55& $10^{ 3}$\\
0.2512& 21.51& 21.51& 11.07&  9.81& 23.24& 23.29& 12.01& 10.54& 23.42& 23.48& 12.15& 10.60& $10^{ 2}$\\
0.3162& 14.25& 14.25&  7.46&  6.55& 15.41& 15.38&  8.10&  7.02& 15.45& 15.45&  8.17&  7.01& $10^{ 2}$\\
\botrule
\end{tabular}
\end{table} 
 
\begin{table}[tbh]
\caption{Neutrino flux ($\rm m^{-2}sec^{-1}sr^{-1}GeV^{-1}$) for $-0.5 \ge \cos\theta_z > -0.6$}
\begin{tabular}{c | r r r r| r r r r| r r r r | c }
\toprule
\multicolumn{1}{c}{}&\multicolumn{4}{|c}{Kamioka}&\multicolumn{4}{|c}{Sudbury }&\multicolumn{4}{|c}{Gran Sasso}&\multicolumn{1}{|c}{}\\
$E_\nu$ (GeV)&$\nu_\mu$~~&$\bar\nu_\mu$~~&$\nu_e$~~&$\bar\nu_e$~~&$\nu_\mu$~~&$\bar\nu_\mu$~~&$\nu_e$~~&$\bar\nu_e$~~&$\nu_\mu$~~&$\bar\nu_\mu$~~&$\nu_e$~~&$\bar\nu_e$~~&$Norm$~ \\
\toprule
0.1000&  7.89&  7.89&  3.83&  3.62&  6.84&  6.93&  3.35&  3.20&  7.88&  7.99&  3.90&  3.66& $10^{ 3}$\\
0.1259&  6.10&  6.10&  2.99&  2.80&  5.35&  5.40&  2.64&  2.48&  6.13&  6.21&  3.05&  2.83& $10^{ 3}$\\
0.1585&  4.59&  4.59&  2.26&  2.08&  4.06&  4.08&  2.00&  1.86&  4.65&  4.68&  2.32&  2.11& $10^{ 3}$\\
0.1995&  3.31&  3.31&  1.65&  1.49&  2.96&  2.95&  1.47&  1.34&  3.36&  3.37&  1.69&  1.51& $10^{ 3}$\\
0.2512&  2.28&  2.28&  1.16&  1.03&  2.07&  2.06&  1.04&  0.93&  2.34&  2.33&  1.19&  1.05& $10^{ 3}$\\
0.3162& 15.22& 15.22&  7.91&  6.90& 13.95& 13.85&  7.13&  6.31& 15.61& 15.58&  8.10&  7.00& $10^{ 2}$\\
\botrule
\end{tabular}
\end{table} 
 
\begin{table}[tbh]
\caption{Neutrino flux ($\rm m^{-2}sec^{-1}sr^{-1}GeV^{-1}$) for $-0.6 \ge \cos\theta_z > -0.7$}
\begin{tabular}{c | r r r r| r r r r| r r r r | c }
\toprule
\multicolumn{1}{c}{}&\multicolumn{4}{|c}{Kamioka}&\multicolumn{4}{|c}{Sudbury }&\multicolumn{4}{|c}{Gran Sasso}&\multicolumn{1}{|c}{}\\
$E_\nu$ (GeV)&$\nu_\mu$~~&$\bar\nu_\mu$~~&$\nu_e$~~&$\bar\nu_e$~~&$\nu_\mu$~~&$\bar\nu_\mu$~~&$\nu_e$~~&$\bar\nu_e$~~&$\nu_\mu$~~&$\bar\nu_\mu$~~&$\nu_e$~~&$\bar\nu_e$~~&$Norm$~ \\
\toprule
0.1000&  8.31&  8.31&  4.07&  3.82&  5.63&  5.71&  2.73&  2.70&  7.40&  7.53&  3.64&  3.49& $10^{ 3}$\\
0.1259&  6.52&  6.52&  3.20&  2.97&  4.52&  4.55&  2.19&  2.14&  5.89&  5.96&  2.90&  2.74& $10^{ 3}$\\
0.1585&  4.94&  4.94&  2.44&  2.21&  3.52&  3.52&  1.70&  1.63&  4.54&  4.56&  2.23&  2.06& $10^{ 3}$\\
0.1995&  3.57&  3.57&  1.78&  1.59&  2.62&  2.61&  1.27&  1.20&  3.33&  3.34&  1.65&  1.50& $10^{ 3}$\\
0.2512&  2.47&  2.47&  1.26&  1.10&  1.88&  1.86&  0.92&  0.85&  2.34&  2.34&  1.17&  1.04& $10^{ 3}$\\
0.3162& 16.41& 16.41&  8.46&  7.30& 12.96& 12.82&  6.38&  5.82& 15.81& 15.72&  8.02&  7.02& $10^{ 2}$\\
\botrule
\end{tabular}
\end{table} 
 
\begin{table}[tbh]
\caption{Neutrino flux ($\rm m^{-2}sec^{-1}sr^{-1}GeV^{-1}$) for $-0.7 \ge \cos\theta_z > -0.8$}
\begin{tabular}{c | r r r r| r r r r| r r r r | c }
\toprule
\multicolumn{1}{c}{}&\multicolumn{4}{|c}{Kamioka}&\multicolumn{4}{|c}{Sudbury }&\multicolumn{4}{|c}{Gran Sasso}&\multicolumn{1}{|c}{}\\
$E_\nu$ (GeV)&$\nu_\mu$~~&$\bar\nu_\mu$~~&$\nu_e$~~&$\bar\nu_e$~~&$\nu_\mu$~~&$\bar\nu_\mu$~~&$\nu_e$~~&$\bar\nu_e$~~&$\nu_\mu$~~&$\bar\nu_\mu$~~&$\nu_e$~~&$\bar\nu_e$~~&$Norm$~ \\
\toprule
0.1000&  7.96&  7.96&  3.89&  3.65&  5.33&  5.39&  2.57&  2.58&  6.59&  6.69&  3.24&  3.11& $10^{ 3}$\\
0.1259&  6.32&  6.32&  3.11&  2.87&  4.33&  4.36&  2.08&  2.06&  5.31&  5.36&  2.60&  2.47& $10^{ 3}$\\
0.1585&  4.84&  4.84&  2.37&  2.16&  3.41&  3.41&  1.63&  1.58&  4.13&  4.15&  2.01&  1.88& $10^{ 3}$\\
0.1995&  3.53&  3.53&  1.75&  1.56&  2.57&  2.56&  1.23&  1.17&  3.07&  3.08&  1.50&  1.37& $10^{ 3}$\\
0.2512&  2.45&  2.45&  1.23&  1.08&  1.86&  1.84&  0.89&  0.83&  2.17&  2.17&  1.07&  0.96& $10^{ 3}$\\
0.3162& 16.33& 16.33&  8.29&  7.17& 12.92& 12.73&  6.23&  5.71& 14.78& 14.68&  7.34&  6.49& $10^{ 2}$\\
\botrule
\end{tabular}
\end{table}
 
\begin{table}[tbh]
\caption{Neutrino flux ($\rm m^{-2}sec^{-1}sr^{-1}GeV^{-1}$) for $-0.8 \ge \cos\theta_z >-0.9$}
\begin{tabular}{c | r r r r| r r r r| r r r r | c }
\toprule
\multicolumn{1}{c}{}&\multicolumn{4}{|c}{Kamioka}&\multicolumn{4}{|c}{Sudbury }&\multicolumn{4}{|c}{Gran Sasso}&\multicolumn{1}{|c}{}\\
$E_\nu$ (GeV)&$\nu_\mu$~~&$\bar\nu_\mu$~~&$\nu_e$~~&$\bar\nu_e$~~&$\nu_\mu$~~&$\bar\nu_\mu$~~&$\nu_e$~~&$\bar\nu_e$~~&$\nu_\mu$~~&$\bar\nu_\mu$~~&$\nu_e$~~&$\bar\nu_e$~~&$Norm$~ \\
\toprule
0.1000&  7.04&  7.04&  3.45&  3.25&  6.42&  6.46&  3.15&  3.01&  6.37&  6.43&  3.15&  2.97& $10^{ 3}$\\
0.1259&  5.67&  5.67&  2.78&  2.57&  5.18&  5.20&  2.53&  2.38&  5.15&  5.19&  2.53&  2.36& $10^{ 3}$\\
0.1585&  4.39&  4.39&  2.13&  1.95&  4.03&  4.02&  1.95&  1.80&  4.01&  4.02&  1.95&  1.79& $10^{ 3}$\\
0.1995&  3.23&  3.23&  1.58&  1.42&  2.98&  2.96&  1.44&  1.31&  2.97&  2.96&  1.45&  1.30& $10^{ 3}$\\
0.2512& 22.57& 22.57& 11.16&  9.84& 20.94& 20.78& 10.22&  9.13& 20.92& 20.79& 10.26&  9.11& $10^{ 2}$\\
0.3162& 15.12& 15.12&  7.54&  6.57& 14.12& 13.96&  6.91&  6.07& 14.12& 13.99&  6.95&  6.08& $10^{ 2}$\\
\botrule
\end{tabular}
\end{table} 
 
\begin{table}[tbh]
\caption{Neutrino flux ($\rm m^{-2}sec^{-1}sr^{-1}GeV^{-1}$) for $-0.9 \ge \cos\theta_z \ge -1.0$}
\begin{tabular}{c | r r r r| r r r r| r r r r | c }
\toprule
\multicolumn{1}{c}{}&\multicolumn{4}{|c}{Kamioka}&\multicolumn{4}{|c}{Sudbury }&\multicolumn{4}{|c}{Gran Sasso}&\multicolumn{1}{|c}{}\\
$E_\nu$ (GeV)&$\nu_\mu$~~&$\bar\nu_\mu$~~&$\nu_e$~~&$\bar\nu_e$~~&$\nu_\mu$~~&$\bar\nu_\mu$~~&$\nu_e$~~&$\bar\nu_e$~~&$\nu_\mu$~~&$\bar\nu_\mu$~~&$\nu_e$~~&$\bar\nu_e$~~&$Norm$~ \\
\toprule
0.1000&  6.78&  6.78&  3.30&  3.15&  8.39&  8.47&  4.17&  3.84&  7.42&  7.49&  3.67&  3.44& $10^{ 3}$\\
0.1259&  5.51&  5.51&  2.66&  2.51&  6.76&  6.80&  3.34&  3.03&  6.02&  6.04&  2.96&  2.72& $10^{ 3}$\\
0.1585&  4.30&  4.30&  2.06&  1.91&  5.22&  5.22&  2.55&  2.27&  4.67&  4.67&  2.27&  2.05& $10^{ 3}$\\
0.1995&  3.18&  3.18&  1.53&  1.39&  3.81&  3.79&  1.86&  1.63&  3.44&  3.42&  1.67&  1.48& $10^{ 3}$\\
0.2512& 22.38& 22.38& 10.84&  9.68& 26.28& 26.09& 12.91& 11.13& 23.92& 23.74& 11.68& 10.19& $10^{ 2}$\\
0.3162& 15.05& 15.05&  7.34&  6.44& 17.32& 17.13&  8.55&  7.25& 15.93& 15.75&  7.78&  6.72& $10^{ 2}$\\
\botrule
\end{tabular}
\end{table} 

\end{document}